\documentclass[journal]{IEEEtran}
\usepackage{amsmath,amsfonts,amssymb}
\usepackage{algorithm}
\usepackage{algorithmicx}
\usepackage{array}
\usepackage[caption=false,font=normalsize,labelfont=sf,textfont=sf]{subfig}
\usepackage{textcomp}
\usepackage{stfloats}
\usepackage{url}
\usepackage{verbatim}
\usepackage{graphicx}
\usepackage{cite}
\usepackage{xcolor}
\usepackage{subfig}
\usepackage{float}
\usepackage{multirow}
\usepackage{longtable}
\usepackage{supertabular}
\usepackage{tabularx}
\usepackage{enumitem}
\usepackage{xspace}
\usepackage{colortbl}
\usepackage{siunitx}
\usepackage{algpseudocode}
\usepackage{booktabs} 

\begin{document}

\newcommand{\sysname}{\textsc{Malsight}\xspace}


\title{\sysname: Exploring  Malicious Source Code and Benign Pseudocode for Iterative Binary Malware Summarization}



\author{
    Haolang Lu, Hongrui Peng, Guoshun Nan,~\IEEEmembership{~Member,~IEEE,} Jiaoyang Cui, Cheng Wang, Weifei Jin, \\Songtao Wang, Shengli Pan,~\IEEEmembership{~Member,~IEEE,} Xiaofeng Tao,~\IEEEmembership{~Senior Member,~IEEE}
    \thanks{H. Lu, H. Peng, G. Nan, J. Cui, C. Wang, W. Jin, S. Pan, X. Tao, are with National Engineering Research Center for Mobile Network Technologies, Beijing University of Posts and Telecommunications, Beijing 100876, China(e-mail: lhl\_2507@bupt.edu.cn; phr@bupt.edu.cn; nanguo2021@bupt.edu.cn; cuijiaoyang24@bupt.edu.cn; wang.me@bupt.edu.cn; weifeijin@bupt.edu.cn; psl@bupt.edu.cn; taoxf@bupt.edu.cn).}
    \thanks{S. Wang (wangst@zgclab.edu.cn) is an assistant researcher at Zhongguancun Laboratory, Beijing, China.}
    \thanks{Corresponding author: G. Nan (e-mail: nanguo2021@bupt.edu.cn).}

}

\markboth{Journal of \LaTeX\ Class Files,~Vol.~14, No.~8, August~2021}%
{Shell \MakeLowercase{\textit{et al.}}: A Sample Article Using IEEEtran.cls for IEEE Journals}


\maketitle

\begin{abstract}
Binary malware summarization aims to automatically generate human-readable descriptions of malware behaviors from executable files, facilitating tasks like malware cracking and detection. Previous methods based on Large Language Models (LLMs) have shown great promise. However, they still face significant issues, including poor usability, inaccurate explanations, and incomplete summaries, primarily due to the obscure pseudocode structure and the lack of malware training summaries. Further, calling relationships between functions, which involve the rich interactions within a binary malware, remain largely underexplored. 

To this end, we propose \sysname, a novel code summarization framework that can iteratively generate descriptions of binary malware by exploring malicious source code and benign pseudocode. Specifically, we construct the first malware summary dataset, MalS and MalP, using an LLM and manually refine this dataset with human effort. At the training stage, we tune our proposed MalT5, a novel LLM-based code model, on the MalS and benign pseudocode datasets. Then, at the test stage, we iteratively feed the pseudocode functions into MalT5 to obtain the summary. Such a procedure facilitates the understanding of pseudocode structure and captures the intricate interactions between functions, thereby benefiting summaries' usability, accuracy, and completeness. Additionally, we propose a novel evaluation benchmark, BLEURT-sum, to measure the quality of summaries. Experiments on three datasets show the effectiveness of the proposed \sysname. Notably, our proposed MalT5, with only 0.77B parameters, delivers comparable performance to much larger Code-Llama. 
\end{abstract}


\section{Introduction}

\begin{figure}[ht]
\centering
\includegraphics[width=0.5\textwidth]{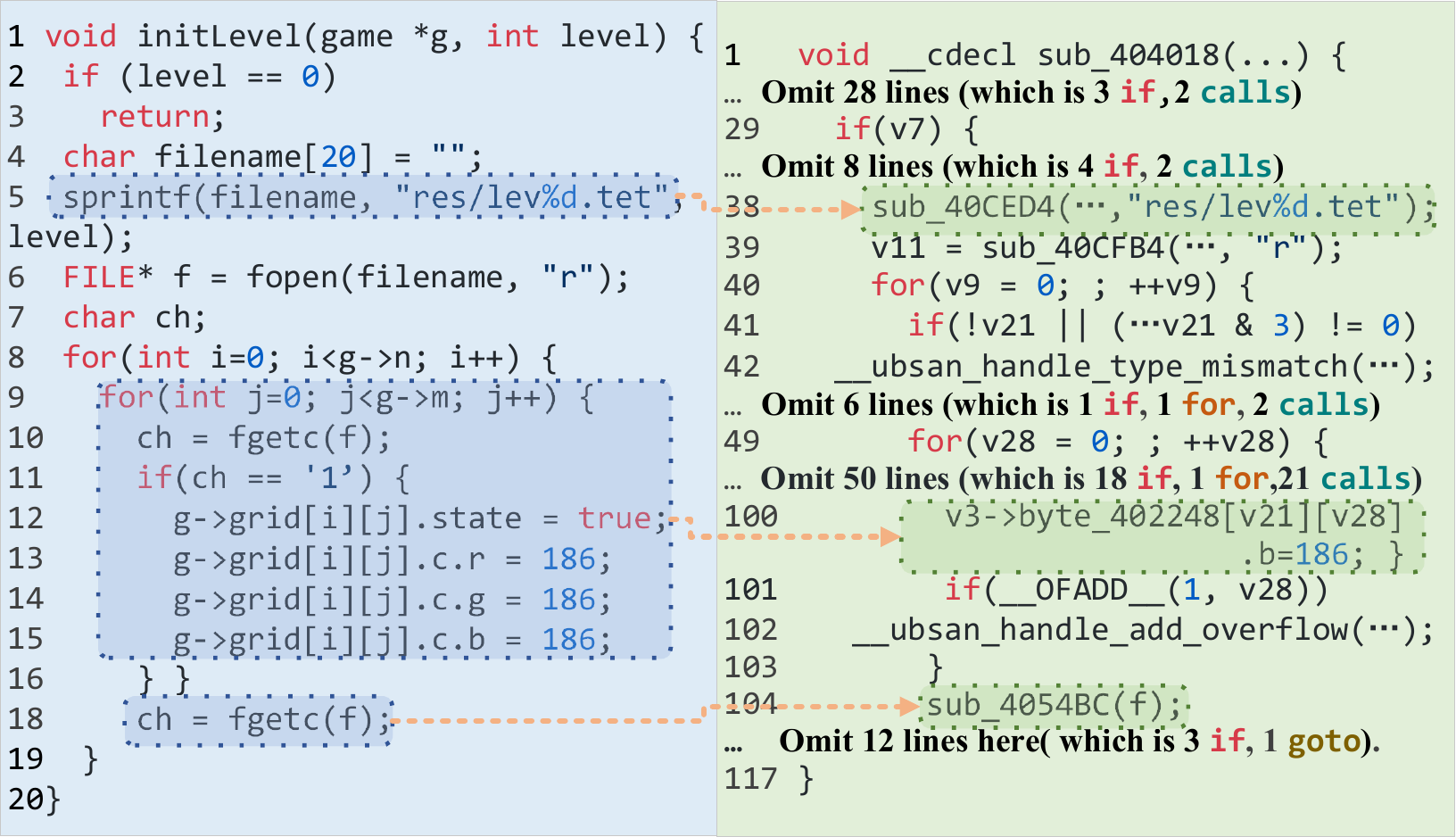}
\caption{The comparison of source code (left) and its pseudocode (right). The pseudocode includes significantly more content and a more complex structure and strips key semantic cues such as function names.} 
\label{fig: intro_src_pseudocode}
\vspace{-5mm}
\end{figure}

\IEEEPARstart{T}{he} AV-TEST Institute~\cite{av-test} recently reported that over 450,000 new malicious files and potentially unwanted applications are registered daily, showing a high demand for malware understanding. Binary malware summarization~\cite{shang2024far} is a reverse engineering~\cite{5941966} task that aims to generate concise human-readable descriptions of binary executable malicious files automatically. The summarization provides security analysts with a quick understanding of the malware's functionality and patterns when source code is unavailable, thereby benefiting a wide range of applications such as malware cracking~\cite{app12178482}~\cite{BEAMAN2021102490}~\cite{10.5555/3620237.3620579}, malware family classification~\cite{GIBERT2020101873}~\cite{10.1145/3319535.3363257}, binary code similarity detection~\cite{280046}~\cite{yu2020order}~\cite{9678518}, and large-scale malware behavior analysis~\cite{220231}~\cite{cornelissen2009systematic}~\cite{9829724}.    

Existing reverse engineering tools, such as IDA~\cite{IDApro} and Ghidra~~\cite{ghidra}, can decompile executables into higher-level C-like pseudocode while still lacking easy-to-understand semantics information. 
While tools such as IDA Pro and Ghidra have significantly advanced binary decompilation by transforming machine code into higher-level pseudocode, they still exhibit critical limitations in handling stripped binaries and entangled logic structures. Specifically, these tools often fail to recover semantically meaningful variable names, function identifiers, or accurate control-flow structures when binaries are stripped of symbolic information. This deficiency leads to pseudocode that is not only semantically impoverished but also structurally convoluted, with deeply nested branches and opaque function calls (e.g., sub\_404018 instead of initLevel), as demonstrated in Fig. 1. Existing small models, although lightweight, often suffer from insufficient capability to capture the complex semantics and entangled logic in stripped malware pseudocode, leading to incomplete or inaccurate summaries.
Consequently, a line of efforts attempts to generate human-readable summaries based on pseudocode. Early studies rely on manual parsing or rule-based summary generation~\cite{10.1145/1858996.1859006}~\cite{10.1145/2597008.2597149}. Recent large language models (LLMs)~\cite{wang2025comprehensive}, such as BinT5~\cite{alkaswan2023extending}, HexT5~\cite{10298504}, CodeGen~\cite{nijkamp2023codegen}, and WizardCoder~\cite{luo2023wizardcoder}, have shown great potential to produce more informative summaries. However, these data-driven approaches still face critical issues, including poor usability, inaccurate explanations, and inaccurate explanations~\cite{shang2024far}. Fig.~\ref{fig: intro_src_pseudocode} shows the underlying reasons for the above issues by comparing the source code of the function ``initLevel'' to the corresponding pseudocode. We observe that the pseudocode presents 1) significantly more content that increases from 20 lines in source to 117 lines in the pseudocode, 2) a more complex and obscure structure with multi-level nesting and entangled logic. The pseudocode involves 29 more calls and 29 more \textit{if} statements compared to the source code at the left, and 3) stripping key semantic cues such as variable names and function names. For example, the function ``initLevel'' in the source code is transferred to symbol ``sub\_404018''. 

\IEEEpubidadjcol

To address the above challenges, we present \sysname 
\footnote{We will release \sysname's source code to contribute to the community.}
, a novel binary malware summarization framework that can iteratively generate descriptions of executable malware by exploring malicious source code and benign pseudocode. The proposed \sysname involves three key ingredients, including a malware dataset MalS, an LLM-based malware summarization model MalT5, and an evaluation metric BLEURT-sum. We describe the workflow of the proposed framework in four steps.  

\textbf{Constructing MalS}: 
LLM-based summarization models rely heavily on high-quality summaries to capture domain-specific knowledge. 
However, the public malware pseudocode summarization dataset is unavailable so far, and building such a benchmark is quite challenging as it requires substantial human involvement for accurate summarization, due to the challenges outlined in Fig.~\ref{fig: intro_src_pseudocode}. 
To address this, we construct MalS, a large-scale dataset with well-constructed summaries based on malicious C-language source code from GitHub.
The proposed MalS involves nearly 90,000 malware source functions, with 20 types of malware functions. We also constructed a small dataset, MalP, for testing. We detail such a procedure in Section~\ref{subsec: 4.3 Building Malware Datasets}.

\textbf{Training MalT5}: 
We propose MalT5, a model fine-tuned on both the MalS dataset and an existing benign pseudocode summarization dataset~\cite{alkaswan2023extending}. 
The underlying intuition is that the malicious semantic knowledge from malware source code summarization and function patterns from benign pseudocode summarization, which is learned from the above two datasets, respectively, can be transferred to the summarization of malware pseudocode. On the one hand, this also effectively addresses the challenges brought about by the lengthy content and complex and obscure structure of decompiled pseudo-code mentioned in the previous text. On the other hand, by doing so, we can properly mitigate the issue of unavailable malware pseudocode summarization datasets. More details are available in Section~\ref{subsec: 4.4 Code Summary Model}.  

\textbf{Performing Generation:} 
Initially, IDApro generates pseudocode and extracts the call graph from a binary file. 
We then develop an algorithm to convert the call graph into a function list, preserving the original call order. 
This function list is iteratively fed into the MalT5 model to generate summaries, following the hierarchy of callers and callees. 
At the same time, we introduce static and dynamic annotation modules to supplement the lack of pseudocode semantics caused by stripping. In addition, in the above process, we iteratively add such annotations to maintain the semantic links between callers and callees. More details are provided in Section~\ref{subsec: code summary - funclist} and~\ref{subsec: 4.2 annotation}.


\textbf{Conducting Evaluation:} 
Previous work~\cite{mastropaolo2023evaluating} indicated that existing metrics for generation tasks, such as Bilingual Evaluation Understudy (BLEU)~\cite{papineni-etal-2002-bleu}, Metric for Evaluation of Translation with Explicit ORdering (METEOR)~\cite{METEORmetric}, Recall-Oriented Understudy for Gisting Evaluation-Longest Common Subsequence (ROUGE-L)~\cite{lin-2004-rouge}, may not well-fit for evaluation of code summarization task. 
We thus employ BLEURT-sum, which is more sensitive to the quality of the pseudocode summary, thereby benefiting the evaluation in practice. More descriptions are given in Section~\ref{subsec: Evaluation Dataset Construction}.

The contribution of this paper can be summarized as follows.

\begin{itemize}
\item A binary malware summarization Framework.  We introduce \sysname, a novel framework that iteratively generates descriptions of binary malware, addressing challenges like entangled logic and stripped semantics.
\item Large-scale datasets for binary malware summarization. We propose \textit{MalS} and \textit{MalP}, two novel datasets that can be used for the LLM training and testing of an LLM of binary malware summarization. To the best of our knowledge, the two datasets are the first in the field.
\item An LLM-based binary malware summarization model. We propose MalT5, a lightweight (0.7B parameters) LLM tailored for binary malware summarization.
\item An evaluation metric for the task: We present BLEURT-sum, a novel evaluation metric more sensitive to pseudocode summarization quality.
\end{itemize}

\section{Related Work}
\subsection{Malware Analysis Engineering}
The field of malware analysis engineering focuses on analyzing malware's functionality by examining its binaries, typically through static analysis methods that involve observing assembly code or pseudocode~\cite{fortinet_malware_analysis}.

\subsubsection{Binary Decompilation}
Decompilation~\cite{7546501} converts executable files into human-readable pseudocode~\cite{hexrays_decompilation_vs_disassembly}, which is more concise and structured than disassembled assembly code. 
Unlike disassembly, which maps instruction encoding directly to assembly statements, decompilation relies on algorithms and patterns (e.g. R2~\cite{radare2}, IDA~\cite{IDApro}, Ghidra~\cite{ghidra}) and emerging methods using LLMs~\cite{tan2024llm4decompile}. However, pseudocode lacks semantic information, such as function names. 
Decompiled function names are often unreadable (e.g., \textit{sub\_4061C0} in IDA Pro)~\cite{sp2024LenorIndex}, providing limited pieces of information for further analysis.

\subsubsection{Human Static Analysis} 
In static analysis, human experts start analyzing from the function entry point~\cite{microsoftMainCommandLine}, inferring functionality from system Application Programming Interface (API) calls, string information, and pseudocode logic.
Their main challenge is accurately identifying the core function~\cite{190918} among numerous functions and methodically tracing the function call~\cite{45287} process to understand the functionality comprehensively.
To assist in this process, we developed Machine Learning-based (ML-based) \sysname, which optimizes and facilitates binary malware analysis.

\subsection{NLP Technologies}
In the \sysname process, we use Bidirectional Encoder Representation from Transformers (BERT)~\cite{devlin-etal-2019-bert} and Text-to-Text Transfer Transformer (T5)~\cite{2020t5} architecture language models to complete specific tasks. 

\subsubsection{BERT Family} 
BERT, a transformer-based model pre-trained with Masked Language Modeling (MLM) and Next Sentence Prediction (NSP), excels in Sequence Labeling (SL) tasks like Named Entity Recognition (NER). 
CodeBERT~\cite{feng-etal-2020-codebert}, extending BERT, learns code semantics through Code-Conditioned MLM and natural language generation, showing strong performance in code tasks. 
Given its BERT origins, we explore CodeBERT for Sequence Labeling (SL) in pseudocode.

\subsubsection{T5 Family} 
T5~\cite{2020t5} is a sequence-to-sequence Transformer model that unifies NLP tasks into a single framework. 
CodeT5~\cite{wang-etal-2021-codet5}, built on T5, extends its capabilities to code understanding and generation using NL-PL bimodal data with identifier tagging and masked identifier prediction. 
CodeT5+~\cite{wang-etal-2023-codet5} improves on this with instruction tuning and additional pre-training tasks, enhancing performance on code-related tasks. 
Previous works like HexT5~\cite{10298504} and BinT5~\cite{alkaswan2023extending} further demonstrate the potential of T5-based models in binary code summarization for malware analysis.

\subsubsection{Transfer Learning} 
Transfer learning leverages abundant labelled data from a source task to help a model learn general features and supplement insufficient training datasets with similar or related data~\cite{jiang2022transferability}. 
In \sysname, we fine-tune the CodeT5+~\cite{wang-etal-2023-codet5} model to achieve transfer learning from the source code summarization task to the decompiled code summarization task. 

\vspace{-0.2cm}
\subsection{Code Summary Evaluation} \label{subsec: word overlap background}
In code summary model evaluation, NLP text similarity algorithms, including word overlap and word embedding measures, compare generated results with reference sets, offering a cost-effective alternative to human evaluation.

\subsubsection{Words' Overlap Measure} 
Early text similarity measures like BLEU~\cite{papineni-etal-2002-bleu} and ROUGE~\cite{lin-2004-rouge} rely on word n-gram overlap between generated and reference text, with BLEU focusing on precision and ROUGE on recall. 
However, they lack semantic understanding. 
METEOR~\cite{METEORmetric} integrates n-gram overlap and semantic similarity using WordNet, providing additional semantic insight.

Recent work~\cite{haque2022semantic} highlights limitations of words' overlap in code summary tasks. It shows that similar structures may yield high similarity scores despite differing semantics.

\subsubsection{Words' Embedding Measure} 
Word2vec~\cite{mikolov2013efficient} is a static embedding model that represents words as points in a vector space, facilitating the proximity of semantically similar words.
MoverScore~\cite{zhao-etal-2019-moverscore} uses an n-gram optimized Word Mover's Distance (WMD)~\cite{10.5555/3045118.3045221} to measure similarity and employs various embedding models like ELMo~\cite{peters2018deep} and BERT.

BLEURT~\cite{sellam-etal-2020-bleurt} stands out as a versatile metric designed for assessing various natural language generation tasks, which combines the advantages of both Words' Overlap Measure and Words' Embedding Measure.
It achieves this by integrating diverse lexical and semantic-level supervision signals into its pre-training process and leveraging synthetic data based on pre-trained BERT, ensuring its effectiveness and versatility in various evaluation scenarios.
\begin{figure*}
\centering
\includegraphics[width=1.0\textwidth]{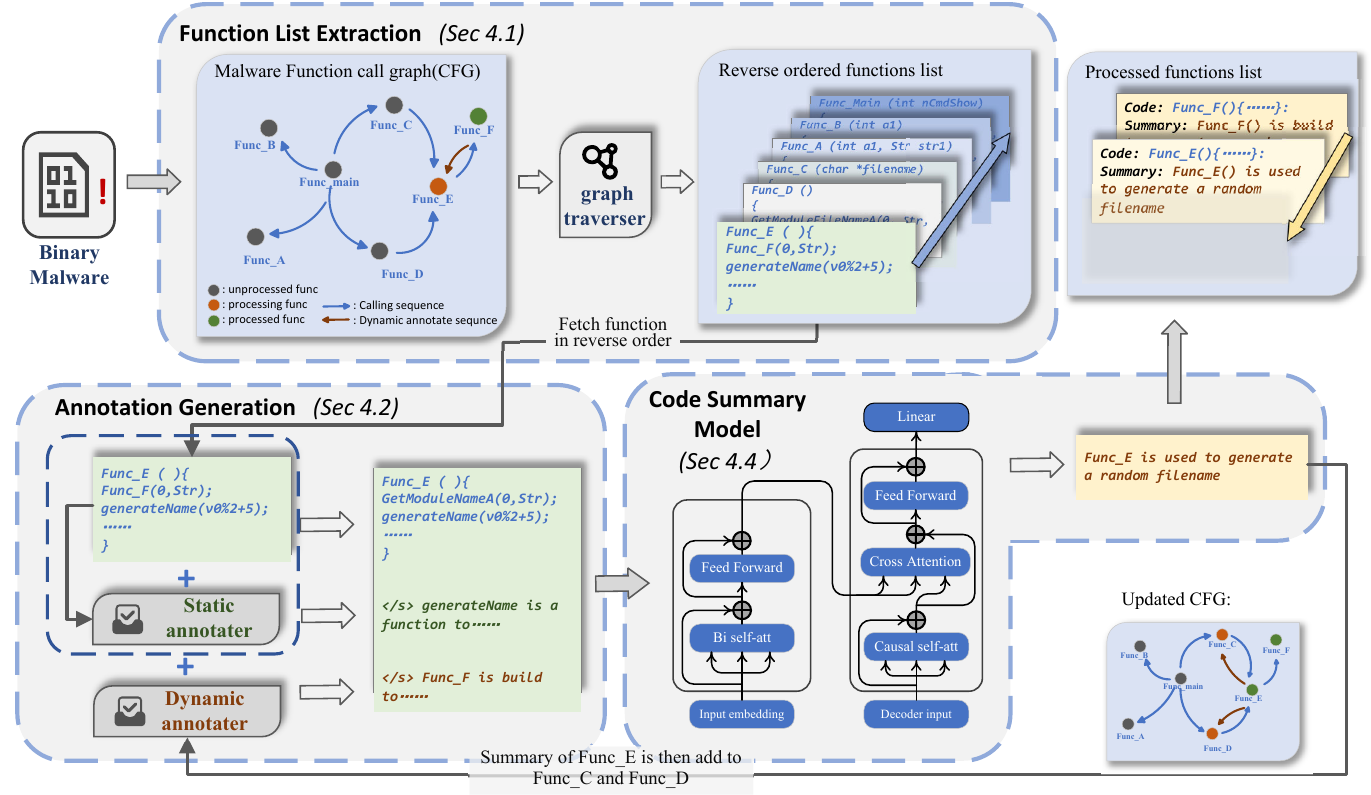}
\caption{Workflow of \sysname. The workflow takes binary malware as input and, after going through graph traversal, annotation generation, and model summarization, iteratively produces function-level code summaries based on the function call hierarchy.} 
\label{fig: System Overview}
\vspace{-0.3cm}
\end{figure*}

\section{Motivation and Overview}
The construction of the code summary framework mainly includes three steps, as shown in Fig.~\ref{fig: System Overview}. During the evaluation phase, we test several evaluation methods and find a reasonable way to build an evaluation model for the code summary task. Simultaneously, our work involves the construction of multiple datasets (for subsequent stages of training and evaluation of MalT5). 

\subsection{Code Summarization Process} \label{subsec: overview-codesummaryprocess}
The code summary task is split into three steps, which are \textit{function list extraction}, \textit{annotation generation}, and \textit{code LLM summary}.

\subsubsection{Function List Extraction} 
Unlike existing code summarization methods focusing solely on a function's internal information, we incorporate the call relationships between functions and treat the entire binary as a processing unit.
As Fig.~\ref{fig: System Overview} shows, it is difficult for the subsequent code summarization model to correctly summarize the functionality of \textit{func\_E} without any information about \textit{func\_F}. (We assume that the function name of \textit{func\_F} has been corrupted.)
Constructing a reverse-ordered function list provides a basis for the subsequent recovery of sub-functions functionality.

\subsubsection{Annotation Generation} 
\sysname will iteratively process the functions in the order of the reverse function list, adding annotations to each function.
In other words, the program sequentially restores functions according to the Function Call Graph (FCG) diagram from the outermost to the innermost and passes function summary results inward.

\subsubsection{Code LLM Summary}
\textit{Fun\_E}(annotated) is then fed into the code summary model for final code summary generation.
In our work, we use transfer learning to adapt the model to both the malware functions' functionality and the decompiled pseudocode's structural features.
The tokenizer splits the code into tokens and embeddings, incorporating a self-attention mechanism into a complete vector in the encoder.
The decoder outputs a fine-tuned prediction based on the code summary and passes it back to the dynamic annotator for subsequent use.

\begin{figure}[htbp]
\centering
\includegraphics[width=0.5\textwidth]{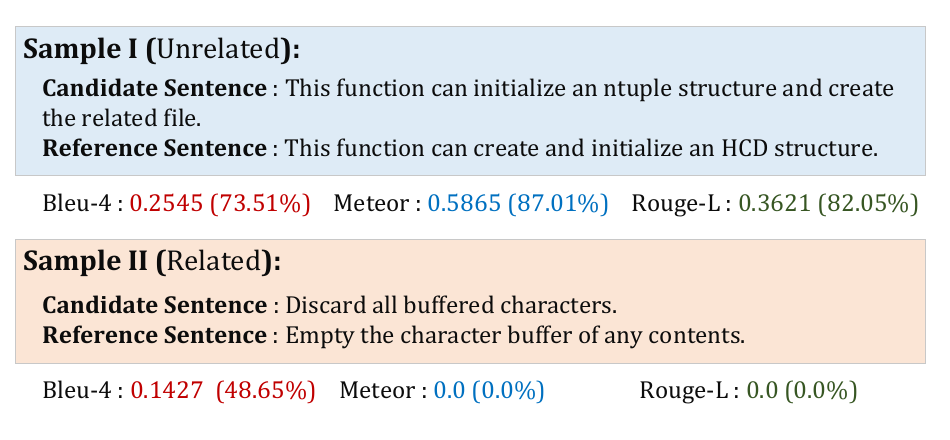}
\caption{A sample of existing evaluation results. Two unrelated pairs of two sentences form Sample I, and two related pairs form Sample II, but the evaluation results contradict expectations.} 
\label{fig: Summary sample}
\vspace{-3.5mm}
\end{figure}

\subsection{Evaluation Method}
We observe that the current popular text similarity measures make it difficult to evaluate the actual quality of the code summary objectively.
Taking Fig.~\ref{fig: Summary sample} as an example, two examples show the evaluation results of BLEU, Meteor, and ROUGE-L on two pairs of real code summaries.
The figure shows that two code summaries without any semantically related results in high evaluation scores (blue-framed), while two semantically similar code abstracts receive low scores (red-framed), demonstrating the shortage of existing methods.
In subsequent work, we propose BLEURT-sum, a BLEURT-based fine-tuning algorithm designed and evaluated to be more suitable for distinguishing available code summaries.

\subsection{Datasets Construction}

For the binary code summary and code summary model evaluation tasks, we build corresponding datasets.

\subsubsection{Dataset For Code Summary Model}
To prevent data shift, training a code summary model requires a large dataset of malware pseudocode summaries. However, malware datasets~\cite{bodmas} are usually available only as compiled binaries, often stripped and structurally complex, making it difficult to create summary datasets without labor-intensive manual efforts. 

Our critical insight is that the code summary model requires two capabilities, \textit{understanding of Malware Functionality} and \textit{adaptability to disassembled pseudocode formats} (including the ability to deal with annotated code). 
Therefore, we separately build datasets from malware source code obtained from GitHub~\cite{259731} and benign pseudocode from the Capybara~\cite{alkaswan2023extending} dataset. They correspond to \textit{MalS} and \textit{BenignC}, respectively, where labels are generated through a carefully designed annotation process that combines large language models with human verification.

We built the \textit{MalP} dataset by compiling and decompiling 20 randomly selected repositories from the above Github malicious code repositories, resulting in 500 functions written in pseudocode for evaluation purpose.

\subsubsection{Dataset For Evaluation Model}
\label{subsec: overview-evaluationdataset}
In the evaluation phase, the evaluation method is used to measure the similarity between the model generation results and the reference results to evaluate the quality of the model generation.
Given two sentences (generation results and the reference results) $S_g$ and $S_r$, most evaluation methods output $Score$ as the evaluation result.
Given the tremendous potential shown by machine learning methods, it is necessary to construct a dataset in the format ${S_g, S_r, Score} ({Score} \in [0, 1] )$ when considering their use.
The challenge lies in obtaining an accurate $Score$ once a dataset of ${S_g, S_r}$ is constructed.
In our subsequent work, we propose a reasonable algorithmic flow for constructing labeled datasets $EvaS$.

\subsubsection{Dataset For Static Annotater}
In the annotation generation process, the static annotater includes a core information extraction module (detail in Section~\ref{subsubsec: Static Annotation}). 
Due to the difficulty in accurately completing the required functions using static methods, we use a machine learning model to complete the sequence labeling(SL) task of the pseudocode. 
By constructing the dataset \textit{CSL}, we have constructed the dataset required for model training and testing.

\begin{table}[htbp]
\caption{The Three Proposed Datasets}
\label{table: datasets}
\resizebox{\columnwidth}{!}{%
\begin{tabular}{ccccc}
\hline
\textbf{}                             & \multicolumn{4}{c}{\textbf{Sets for code summary model}}                                  \\ \hline
\multicolumn{1}{c|}{\textbf{Datasets}} & \textbf{Size(functions)} & \textbf{Code language}          & \textbf{Annotated?}    & \textbf{Usage} \\ \hline
\multicolumn{1}{c|}{\textit{MalS}}    & 89,609                & C                    & No                   & Train phase1         \\
\multicolumn{1}{c|}{\textit{MalP}}    & 500                  & pseudo               & Yes                  & Test                 \\
\multicolumn{1}{c|}{\textit{BenignC}} & 96,835                & pseudo               & Yes                  & Train phase2         \\
\multicolumn{1}{l}{}                  & \multicolumn{1}{l}{} & \multicolumn{1}{l}{} & \multicolumn{1}{l}{} & \multicolumn{1}{l}{} \\ \hline
\multicolumn{1}{c|}{}                 & \multicolumn{4}{c}{\textbf{Sets for annotation extractor model}}                          \\ \hline
\multicolumn{1}{c|}{\textbf{Dataset}}  & \textbf{Size(functions)} & \textbf{Code language}          & \textbf{Anno num(avg)} & \textbf{Usage} \\ \hline
\multicolumn{1}{c|}{\textit{CSL}}   & 95,000                & pseudo               & 3.87                 & Train \& Test        \\
\multicolumn{1}{l}{}                  & \multicolumn{1}{l}{} & \multicolumn{1}{l}{} & \multicolumn{1}{l}{} & \multicolumn{1}{l}{} \\ \hline
\multicolumn{1}{c|}{}                 & \multicolumn{4}{c}{\textbf{Sets for evaluation model}}                                    \\ \hline
\multicolumn{1}{c|}{\textbf{Dataset}}  & \textbf{Size(pairs)}     & \textbf{Pos\textbackslash{}Neg} & \textbf{Length(Avg)}   & \textbf{Usage} \\ \hline
\multicolumn{1}{c|}{\textit{EvaS}}    & 127,510               & 1:1                  & 9.6                  & Train \& Test        \\ 
\end{tabular}%
}
\end{table}

To sum up, we construct three datasets to address the different tasks arising from the various stages of the \sysname framework, as shown in Table~\ref{table: datasets}.


\section{Code Summarization Workflow}

Following our breakdown of the malware code summary task in Fig.~\ref{fig: System Overview}, we construct comprehensive systems incorporating multiple LLMs and algorithms.

\subsection{Function List Extraction} \label{subsec: code summary - funclist}

In the first module of \sysname, we extract a reverse function list from the malware's FCG. 
Since existing methods~\cite{10179482} don't fully ensure accurate FCG extraction, we implement a pluggable module to map the binary $B_{Mal}$ to the FCG digraph $G_M$.
Using an inverse topological traversal algorithm, $REsort$, we generate the reverse order list $L_{G_M}$, where functions are arranged according to their inverse topological order. 

\definecolor{darkblue}{rgb}{0.0, 0.0, 0.75}
\begin{algorithm}
    \caption{REsort}
    \label{alg: Tarjan}
    \begin{algorithmic}[1]
        \Require The FCG graph $G_M$ for a malware binary file.
        \Ensure The reverse topsort list $L_G$ of the function call graph.
        \Function {REsort}{$G_M$}
            \For{each vertex $v$ in $G_M.vertices$}
                \If{$!v.seen$}
                    \State \textbf{new} $G_M'$
                    \State $G_M'.vertices \leftarrow$ \textbf{call} Tarjan($v$, $G$)
                \EndIf
            \EndFor
            \State $G_M' \leftarrow$ \textbf{call} BuildTarGraph($G_M'.vertices$, $G_M$)
            \State $L_{G_M'} \leftarrow$ \textbf{call} RetopSort($G_M'$)
            \State $dist[] \leftarrow$ \textbf{call} Dijkstra($G_M$)

            \State$L_{G_M} \leftarrow []$
            \For{each vertex $v'$ in $L_{G_M'}$}
                \State $mindist \leftarrow \infty$, $idx \leftarrow -1$

                \For{each $ver$ in $v'.subvertex$}
                    \If{$dist[ver] < mindist$}
                        \State $mindist \leftarrow dist[ver]$, $idx \leftarrow ver$
                    \EndIf
                \EndFor
                \State $L_{G_M}$.append(DFS($idx$, $v'$))
            \EndFor
        \EndFunction
    \end{algorithmic}
\end{algorithm}

Algorithm~\ref{alg: Tarjan} shows that equipped with Tarjan, Dijkstra, and DFS, our proposed algorithm is able to address challenging issues such as cyclic calls and partially connected graphs. 
The function list $L_{G_M}$ is processed from outer API calls to the main function to ensure the correct restoration order.
Specifically, $L_{G_M}$ follows the call hierarchy from the outermost to the innermost layers, ensuring that functions called later appear earlier in the list and are processed first in subsequent steps.

\subsection{Annotation Generation } \label{subsec: 4.2 annotation}

\begin{figure}
\centering
\includegraphics[width=0.45\textwidth]{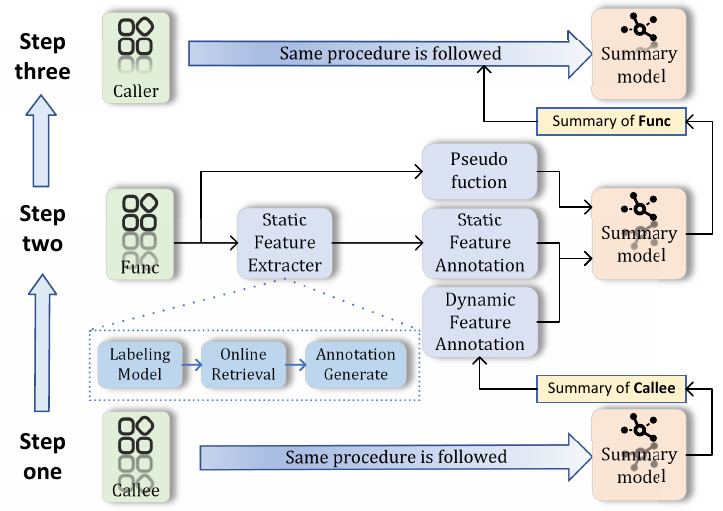}
\caption{Annotation workflow. If the function currently being processed is Func, its dynamic annotation comes from the child function it calls callee, and the static annotation is generated in three steps. Func's code summary is then passed to the Caller as a dynamic annotation.} 
\label{fig: annotation process}
\vspace{-0.3cm}
\end{figure}

In $L_{G_M} = \{f_1, f_2... f_{len(L_{G_M})}\}$, we sequentially add static and dynamic annotations to enrich the information for each function $f_i$. 

As mentioned, attackers often use techniques like stripping to remove symbolic information, leading to significant semantic loss in the pseudocode and hindering code summarization models. 
To address this, we propose using dynamic and static annotations to supplement the missing semantics and enhance pseudocode features. 
This approach, as illustrated in Fig. ~\ref{fig: annotation process}, helps bridge the gap between source code and stripped pseudocode.

\subsubsection{Static Annotation} \label{subsubsec: Static Annotation}
Through our extensive study of stripped malware pseudocode, we identified that critical API calls (e.g., OS APIs) and certain preserved strings offer valuable clues for malware behavior analysis and semantic recovery despite the significant semantic loss. 
Thus, we design a static annotation module to enhance code summaries, which operates in three key steps: sequence labeling, online retrieval, and annotation generation.

\textbf{Sequence labeling model:}
By manually labeling approximately 300,000 tokens within nearly 95,000 functions with the assistance of  lex~\cite{beazley2001ply}, we construct a labeled dataset \textit{CSL} based on \textit{BenignC} to train the sequence labeling model. 

Formally, the function $f_i$ is first sliced into an n-token code sequence by the tokenizer $T$.
The n-token code sequence $s_i = T(f_i) = \{t_0, t_1, ..., t_{n-1}\}$ and the labels of the tokens $L_i = \{l_0, l_1, ..., l_{n-1}\}$ are combined into $d_i = \{(t, l) | t \in s_i, l \in L_i \}$.
Equation~\eqref{equ: CSL_ds_eq} formalizes this dataset.
\begin{equation}
CSL = \{d_0, d_1, ..., d_N\}
\label{equ: CSL_ds_eq}
\end{equation}

Subsequently, we utilize the CodeBERT~\cite{feng-etal-2020-codebert} model for training the SL task using this \textit{CSL}. 
As shown in equation~\eqref{equ: B_eq} and~\eqref{equ: C_eq}, $B$ is the CodeBERT base model, and $C$ is the classifier head. 
As a sequence $s_i$ is inputted, the complete model outputs the predicted labels $y_i$ of its tokens $o_i$.

\vspace{-0.2cm}
\begin{equation}
B(s_i) = \{o_0, o_1,..., o_{n-1} \}
\label{equ: B_eq}
\vspace{-0.1cm}
\end{equation}
\begin{equation}
C(o_i) = y_i
\label{equ: C_eq}
\end{equation}

Further, the target function is formalized in equation~\eqref{equ: LF_eq}, where $l_{i_{c}}$ is the truth label and $y_{i_{c}}$ is the Softmax probability for the $c^{th}$ class.
\vspace{-0.1cm}
\begin{equation}
LF = -\sum\limits_{c=0}^{2} l_{i_{c}}\log{y_{i_{c}}}
\label{equ: LF_eq}
\end{equation}

By optimizing $LF$, $B$ and $C$ are trained simultaneously. 
This necessitates that the model effectively classifies code-tokens to accurately label the key API calls and special strings within the pseudocode.

\textbf{Label to annotation:}


Once we obtain the key API calls and special string literals from the stripped pseudocode, we utilize GitHub Code Search~\cite{github_code_search} to retrieve relevant contextual information from public GitHub repositories.

In practice, we construct a query based on the extracted identifiers and submit it to the code search engine. To ensure relevance while maintaining efficiency, we retain only the top three retrieved blocks for each query. This decision is motivated by the observation that GitHub Code Search sorts results by estimated relevance according to their ranking algorithm~\cite{techBehindCodeSearch}, and that the top-ranked blocks typically contain the most pertinent information.

We conducted random sampling and manual inspection across the retrieved results. The analysis shows that approximately 54.8\% of the retrieved function contexts contain comments explicitly describing the function's intended functionality. Among the remaining functions, nearly 90\% still provide valuable contextual signals, such as meaningful parameter names, semantically related helper functions, control flow hints, and call dependencies. Only a small minority of cases yield irrelevant or unusable retrievals, often due to noise in the search index or ambiguous identifiers. After retrieval, the code snippets undergo a filtering and preprocessing step. Specifically, we discard results that are excessively short, semantically incomplete, or duplicate the input pseudocode without adding new context. The remaining high-quality snippets are then incrementally injected into the prompt of a generic, instruction-following large language model to generate static natural language annotations for each function, denoted by $AnnoS(f_i)$. This retrieval-augmented annotation strategy significantly enriches the available context for static analysis and serves as an important intermediate process in our overall framework.

\subsubsection{Dynamic Annotation}
As shown in Fig.~\ref{fig: annotation process} (the blue parts represent the steps in which the annotation was added), the summary of the callee is provided to the caller as a complement to the semantic information, which we define as dynamic annotation. 
This is consistent with the actual analysis flow of binary malware analysis by reverse workers, i.e., analyzing the call relationship from the inner layer of the FCG diagram to the outer layer (corresponding function list generated in Section~\ref{subsec: code summary - funclist}) and summarizing the function from the outside in (corresponding to the passing of dynamic annotation).
In this way, we can make full use of the dynamic behavior characteristics implied by the call relationships between functions in the pseudocode, denoted by $AnnoD(f_i)$.
The function $f_i = f_i\oplus AnnoS(f_i) \oplus AnnoD(f_i)$ will be passed into the code summary model for later processing.

\subsection{Building Malware Datasets} \label{subsec: 4.3 Building Malware Datasets}
In the traditional scheme of building datasets for decompiled code, the datasets are built at the function level~\cite{ye-etal-2023-cp}.
The source function $f_n^{Source}$ is compiled and linked with other modules to generate an executable file, as $f_n^{Bin}$, and then decompiled to obtain the pseudocode form $f_n^{pseudo}$ of the corresponding function. Equation~\eqref{equ: ideal_dataset} formalizes this dataset, where $SUM()$ is the extraction method for the code summary.
\begin{equation}
Set_{ideal} = \{(f_n^{pse},f_n^{sum})|f_n^{sum}=SUM(f_n^{pse})\}
\label{equ: ideal_dataset}
\end{equation}

Out of 2,289 GitHub repositories that are determined to be malware, we extract close to 30K functions. 
We filter out functions that are repetitive, shorter than five lines, or have formatting issues, resulting in a dataset of 89,609 functions.
(The lack of strict filtering may lead to overlap between the train and test sets, consequently yielding inflated results.)

Unlike well-maintained open-source projects, most extracted functions lack comments for use as code summaries. 
However, the source code's rich semantics allowed us to generate summaries using a well-designed prompt with LLM, and engaged five human experts to assist in reviewing and refining the generated summaries, resulting in the \textit{MalS} dataset in equation~\eqref{equ: MalS_dataset}.
We built the \textit{MalP} dataset with 500 functions, selected from 20 randomly chosen repositories. 
Its smaller size is due to the difficulty of manually compiling and decompiling poorly maintained malware repositories, from which we picked functions with strong functionality characteristics.
\vspace{-0.1cm}
\begin{equation}
Set_{MalS} = \{(f_n^{Sou},f_n^{sum})|f_n^{sum}=SUM(f_n^{Sou})\}
\label{equ: MalS_dataset}
\end{equation}
\begin{equation}
Set_{MalP} = \{(f_n^{pse},f_n^{sum})|f_n^{sum}=SUM(f_n^{Sou})\}
\label{equ: MalP_dataset}
\end{equation}

Since \textit{MalS} is a source code training set, we need an additional dataset to help our model understand stripped pseudocode features.
The Capybara dataset provided by BinT5~\cite{alkaswan2023extending} is a benign pseudocode dataset.
We process the Capybara dataset ($Set_{Capybara}$) using our system's static annotation module ($ANNS()$), transforming it into \textit{BenignC}.
\begin{equation}
Set_{Capybara} = \{(f_n^{pse},f_n^{sum})|f_n^{sum}=SUM(f_n^{Sou})\}
\label{equ: Capybara_dataset}
\end{equation}
\begin{equation}
Set_{benignC} = \{(f_n^{ann},f_n^{sum})|f_n^{ann}=ANNS(f_n^{pse})\}
\label{equ: benign_dataset}
\end{equation}

When training the code summary model, we use \textit{MalS} and \textit{BenignC} to complete the transfer learning tuning process, and $Set_{MalP}$ to test during the evaluation phase.

\begin{figure}
\centering
\includegraphics[width=0.5\textwidth]{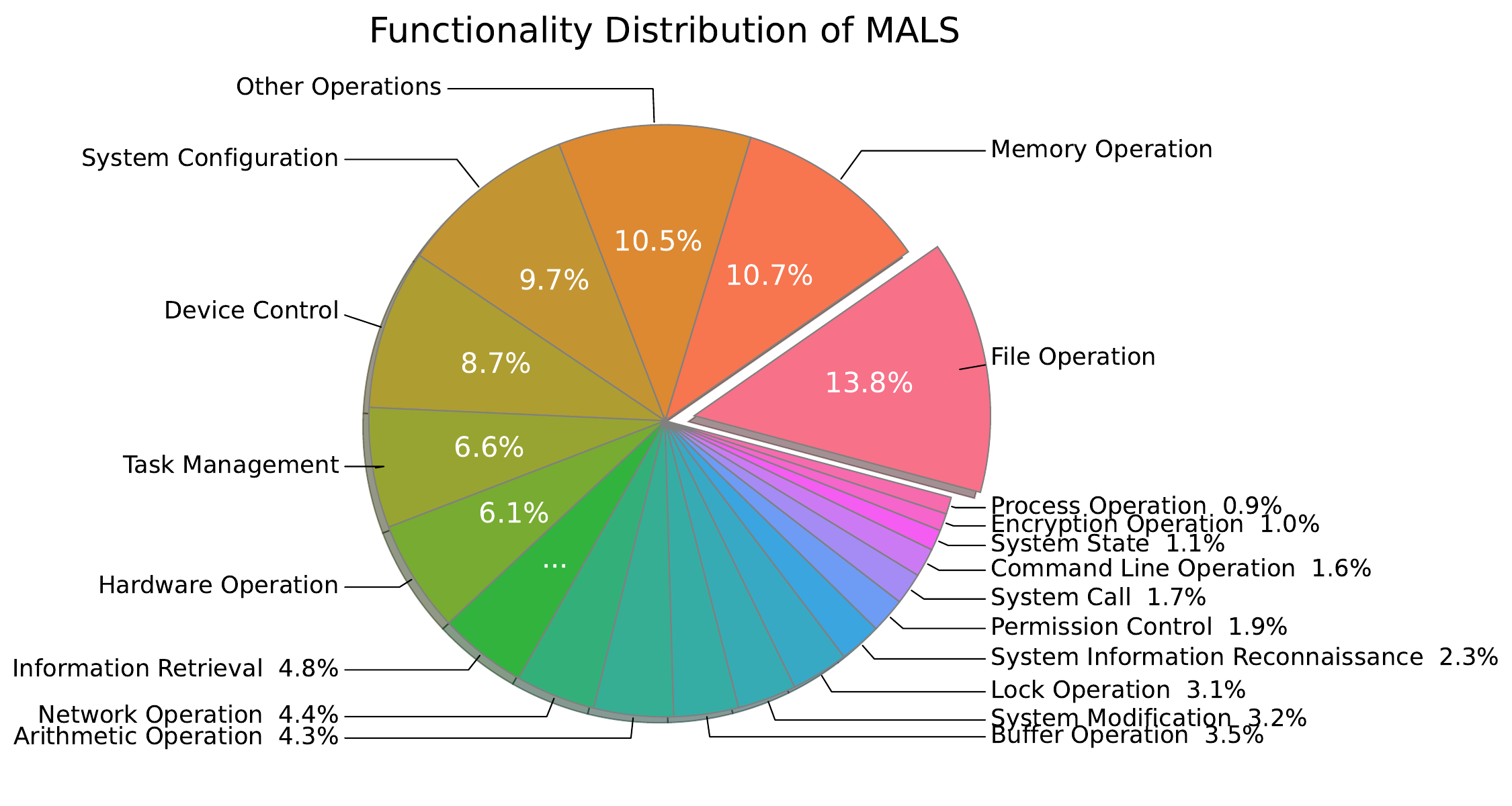}
\caption{Topic distribution of code summary. The results of topic analysis for ${Set_{MalS}}$ indicate that its distribution conforms to the functional distribution of malware functions.} 
\label{fig: dataset_distribution}
\vspace{-0.3cm}
\end{figure}

Additionally, 25\% samples from \textit{MalS} were randomly extracted for thematic analysis of function functionality, as shown in Fig.~\ref{fig: dataset_distribution}.
Our analysis results confirm that \textit{MalS} and \textit{BenignC} have significantly different theme distributions. 
Among a large number of security-related functions in the former, there are an unignorable number of codes for lock, permissions, and encryption, while in the latter, there are a large number of codes related to game logic and driver calls which irrelevant to malware.

\subsection{Code Summary Model} \label{subsec: 4.4 Code Summary Model}
In \sysname, we fine-tune the CodeT5+ model to enable transfer learning from source code summarization to decompiled pseudocode summarization. 
To achieve this, we design two distinct phases of fine-tuning to facilitate smooth transfers that accommodate variations in data characteristics and distribution biases across different datasets.

In \sysname, we leverage transfer learning to shift the model from source code summarization to decompiled pseudocode summarization. 
This is achieved by utilizing two datasets: \textit{MalS} (source code dataset) and \textit{BenignC} (pseudocode dataset), aiming to approximate the ideal training effect on an unachievable dataset $Set_{ideal}$. 

First, the model is trained on \textit{MalS} for semantic features of the malware source code. 
For each sample $(f_n^{Sou},f_n^{sum})$ in \textit{MalS}, where $f_n^{Sou}$ represents a source code snippet and $f_n^{sum}$ represents the corresponding summary, the model minimizes the loss function as equation~\eqref{equ: mals_train}. 
Here, $M_\theta$ denotes the model parameters, and $\mathcal{L}$ represents the loss function (e.g., cross-entropy loss). 
The output of this stage is the encoded feature representation of the source code $h_{MalS} = \text{Encoder}_\theta(x_i)$.

\begin{equation}
L_{MalS} = \frac{1}{n_{MalS}} \sum_{i=1}^{n_{MalS}} \mathcal{L}(M_\theta(f_n^{Sou}), f_n^{sum})
\label{equ: mals_train}
\end{equation}

Next, the model undergoes transfer learning using \textit{BenignC} (pseudocode dataset) to further refine its capabilities. For each sample $(f_i^{ann}, f_n^{sum})$ in \textit{BenignC}, where $f_i^{ann}$ is the pseudocode sequence and $f_n^{sum}$ is the annotation, the data processing pipeline is shown in equation~\eqref{equ: bimodal_seq}.

\begin{equation}
f_i^{ann} \to \{fc_0^{ann}, \dots, fc_{p-1}^{ann}, f_{sep}, fc_0^{ann}, \dots, fa_{q-1}\}
\label{equ: bimodal_seq}
\end{equation}

The input pseudocode and annotation are concatenated into a bimodal sequence where $fc_p$ represents tokens from the pseudocode, $fa_q$ represents tokens from the annotation, and $f_{sep}$ is a special separator token.

The T5 model's encoder processes this sequence of pseudocode and annotations, generating an encoded representation $h_{benignC} = \text{Encoder}_\theta(f_i^{ann})$. This representation serves as the basis for generating the corresponding code summary. To optimize the model’s performance, we minimize the loss function defined in equation~\eqref{equ: ben_train}, where $M_\theta$ represents the parameters of the model, $f_i^{ann}$ is the input pseudocode sequence, and $f_i^{sum}$ is the target summary.

\begin{equation}
L_{benignC} = \frac{1}{n_{benignC}} \sum_{i=1}^{n_{benignC}} \mathcal{L}(M_\theta(f_i^{ann}), f_i^{sum})
\label{equ: ben_train}
\end{equation}

This loss function captures the difference between the predicted and target summaries across the entire dataset.  The model leverages transfer learning by training on both \textit{MalS} (source code) and \textit{BenignC} (pseudocode), allowing it to learn generalized features from source code and apply them to pseudocode summarization.  By facilitating knowledge transfer between these datasets, the T5 encoder ensures consistency and improves its ability to handle different data modalities, enhancing performance on both source code and pseudocode tasks.

\begin{equation}
L_{total} = \alpha L_{MalS} + \beta L_{benignC}
\label{equ: train_pro}
\end{equation}

The overall objective loss for the transfer learning process is defined in equation~\eqref{equ: train_pro}.
Specifically, $\alpha$ and $\beta$ are balancing factors between the two learning objectives. 
This method allows the model to adapt to the semantic structure of pseudocode and generate high-quality code summaries.
\section{Evaluation Method}

As mentioned in Section~\ref{subsec: overview-evaluationdataset}, the evaluation algorithm for code summary tasks takes the reference as input and \textit{distinguish between valid and invalid generated results.}\label{Eva model insight}
This section introduces our exploration of EvaS and BLEURT-sum, respectively.

\subsection{Evaluation Dataset Construction}\label{subsec: Evaluation Dataset Construction}
Utilizing the code summary from the \textit{MalS} dataset as a foundation for our research, we curate a positive and negative sample pair for the tuning of our new evaluation model BLEURT-sum.
The positive sample consisted of two code summary result sentences sharing the same meaning is initialized as $\{S_g, S_r, 1\}$, while the negative sample comprised two randomly different sentences and is initialized as  $\{S_g, S_r, 0\}$. 
$S_g$ and $S_r$ represent generated statements and reference statements, respectively.

To build the dataset in $\{S_g, S_r, Score\} ({Score} \in [0, 1] )$  format, one approach is to build an algorithm, $Score = GenSim(S_g, S_r,0 or 1)$, which automatically generate the $Score$ labels needed for model training.
The key task is constructing a reliable $GenSim()$ function.
When $S_g == SUM(f_n)$, the $Score$ can be expressed using equation~\eqref{equ: Score1}.
This accounts for the fact that the SUM() function may produce different outputs for the same input, as it is a non-deterministic function.

{\small
    \begin{equation}
    \hspace{-0.4cm}
    S_r \! = \! \text{SUM}(f_n)
    \! \implies \!
    Score \! =  \! \begin{cases}
    1 \! & \! \text{if } S_g = \text{SUM}(f_n) \\
    0 \! & \! \text{if } S_g = \text{SUM}(\neg f_{n})
    \end{cases}
    \label{equ: Score1}
    \end{equation}
}

When constructing the $\{S_g, S_r\}$ pairs, sentence structure dependencies are not considered, meaning that the score calculated in equation~\eqref{equ: Score1} is based purely on semantic information. 
To enhance this, we propose incorporating both sentence structure and semantics into the evaluation process. 
Our approach seeks to determine the appropriate balance between these two aspects in sentence similarity evaluation.

Taking the original 0,1 tag as the semantic feature, we further extract the static feature to get $s_f$.
Assuming that the proportion of semantic information is $p$, the structural feature calculation function is $Struc()$, 
the following equations formalize this Score construction method.

{\small
    \begin{gather}
    s_f = Struc(S_g,S_r)   \\
    Score \! = \! \begin{cases}
    p + (1-p)*s_f \! & \! \text{if} S_g = \! \text{SUM}(f_n) \\
    (1-p)*s_f  \! & \! \text{if} S_g = \! \text{SUM}(f_{\neg n})
    \end{cases}
    \label{equ: Score2_2}
    \end{gather}
}

In the implementation, we utilize the arithmetic average of BLEU, ROUGE-L, and METEOR as the metric for $Struc()$ (they have been normalized to the zero-one interval using a uniform probability distribution, respectively).
Then we solve the minimum square error and get the semantic and structural ratio close to 1:4, which makes $p = 1/5$.

\subsection{Evaluation Model}
Based on our insights of the evaluation model, it should be able to combine both structural and semantic features and give an evaluation score for the specific task of code summary.

BLEURT is based on BERT and adds additional pre-training steps between pre-training and fine-tuning to the synthesized data. Synthetic data is generated by perturbing sentence pairs $<z,\widetilde z>$, where $z$ and $\widetilde z$ are randomly selected sentence pairs. 
In the additional pre-training of BLEURT, a series of pre-training signals $(\tau_1,\tau_2,... \tau_9)$ to align the model with the desired result. BLEURT uses the sentence pair scores of BLEU, ROUGE, and BERTScore as signals $\tau_1$ to $\tau_3$, and uses the back translation processing sentence pairs to generate $\tau_4$ to $\tau_7$. 

The rich training signals ensure the universality of BLEURT and the comprehensiveness of the evaluation angle. 
We then further fine-tune the BLEURT model on EvaS dataset.
First, the generated text and the reference text are input together into the model and the vector is generated as shown in equation~\eqref{equ: bleurtembedding}.

\begin{equation}
v_{[CLS]},v_{S_{g1}},...,v_{S_{gn}},...,v_{S_{rn}} = BLEURT(S_g, S_r)
\label{equ: bleurtembedding}
\end{equation}

Further, the model uses the CLS vector to add antecedents to obtain the scores predicted by the model, as shown in equation~\eqref{equ: bleurtgenscore}.
\begin{equation}
\hat{Score}=f(S_g, S_r)=W\tilde{v}_{[CLS]}+b
\label{equ: bleurtgenscore}
\end{equation}

Finally, the model starts to complete the loss calculation based on the loss and then carries out the gradient descent (equation~\eqref{equ: bleurtloss}).
\begin{equation}
loss=\frac1N\sum\limits_{n=1}^N\left|\left|Score-\hat{Score}\right|\right|^2
\label{equ: bleurtloss}
\end{equation}

In summary, existing word overlap methods often inflate scores due to common keywords like "return" or "initialize", which may be incremental. 
Meanwhile, the BLEURT-sum improves performance by considering both word overlap and additional dimensions.
\section{Experiments}

\begin{table*}[hbtp]
\centering
\caption{Comparisons with baseline work}
\label{table: comparision} 
\resizebox{\textwidth}{!}{%
\begin{tabular}{c|cccc|ccc}
            & BLEURT-sum & BLEU & ROUGE-L & METEOR & AVG time (function)
& AVG summary length
& BLEURT-sum variance \\ \hline
BinT5& 21.18 & 1.92 & 9.45 & 3.51 & \underline{0.16}& 6.81  
& 201.11 \\
HexT5& 27.67& 2.71& 11.23& 3.22& \textbf{0.13}& 7.11& 216.81\\ 
WizardCoder-15B& 53.43 & 7.75 & 23.16 & 13.71 & 2.44 
& 32.20 
& 303.03 \\
Code Llama-7b       & \underline{56.00} & \underline{8.52} & \underline{24.55} & \underline{14.95} & 2.24 
& 31.32
& 290.16 \\
Code T5+& 17.18& 1.74& 4.17& 2.54& 0.1743& 7.2869& \underline{161.34}\\
WizardLM-2-7B& 55.81 & 5.33 & 18.46 & 15.61 & 22.07 
& 61.74 
& 265.63 \\
deepseek-llm-7b-chat& 50.88 & 7.07 & 20.08 & 12.70 & 7.41 
& 25.48 
& 306.78\\
\sysname (Ours)& \textbf{62.14} & \textbf{9.80}& \textbf{25.11}& \textbf{16.87} & 2.51& 35.27&\textbf{131.65} \\ \hline
\multicolumn{8}{l}{\footnotesize * AVG time (function) is measured in seconds.}\\
\multicolumn{8}{l}{\footnotesize * AVG summary length is measured in words.}
\end{tabular}%
}
\vspace{-0.5cm}
\end{table*}

Our evaluation experiments are designed to answer the following four questions:
\begin{enumerate}
    \item RQ1 (§~\ref{sec: RQ1}). How does \sysname performance compare across two training phases and mainstream code summarization models? 
    \item RQ2 (§~\ref{sec: RQ2}). How do module combinations and different stripping scenarios affect \sysname's performance, particularly the annotator module?
    \item RQ3 (§~\ref{sec: RQ3}). How does the new BLEURT-sum evaluation method compare to existing code summarization evaluation metrics? 
    \item RQ4 (§~\ref{sec: RQ4}). How does \sysname perform when applied to real-world malicious software, and how do assessments from human reverse engineers validate its usability?
\end{enumerate}

\subsection{Experimental Setup}

Our experiments run on the Ubuntu 20.04 system, equipped with one Intel Xeon Silver 4210 CPU 2.20 GHz, and two NVIDIA A40 GPUs with 125 GB RAM. Binary file processing tools include Radare2, IDA Pro v7.5, GCC v9.4.0, and GNU Make v4.2.1. Our programming language is Python v3.8.13, with transformers v4.16.2, torch v2.1.2+cu121.

\subsection{RQ1. Performance Test} \label{sec: RQ1}
\subsubsection{Code Summary Performance}
During training, the model undergoes two phases. The first phase involves fine-tuning malware source code (\textit{MalS}) to learn malicious behavioral features. The second phase focuses on generating high-quality code summaries (\textit{BenignC}) by incorporating semantic information from pseudocode and annotations.
In both phases, we split the dataset into training, cross-validation, and test sets with a ratio of 7:2:1. We then evaluate the model at each stage to ensure its effectiveness.
The result is shown in Table~\ref{table: twophase}. 

\begin{table}[!htbp]
\caption{Score During Two Train Phases}
\resizebox{\columnwidth}{!}{%
\begin{tabular}{l|llll}
       & BLEURT-sum & BLEU & ROUGE-L & METEOR \\ \hline
Phase1 & 74.17      & 18.06 & 38.56  & 24.69   \\
Phase2 & 72.74      & 22.61 & 41.19  & 25.37   \\
CP-BCS(Baseline) & 17.78 & 21.50 & 16.89  & 11.92 \\ \hline
\end{tabular}%
}
\label{table: twophase}
\end{table}

\begin{table*}[htbp]
\centering
\caption{Performance comparisons on various levels of stripping}
\resizebox{\textwidth}{!}{%
\begin{tabular}{c|cccc|ccc}
Stripping Levels  & BLEURT-sum & BLEU & ROUGE-L & METEOR & AVG time (function)
& AVG summary length & BLEURT-sum variance 
\\ \hline
Not-Stripped   & 66.29  & 10.95  & 26.97  & 18.80 & 2.28 
& 32.85 & 111.39 
\\
Not-Stripped w/o Annotation    & 60.05 & 7.65 & 22.76 & 14.93 & 2.01 
& 29.12 & 143.62 
\\
Demi-Stripped   & 64.01 & 10.42 & 26.03 & 17.62 & 2.44 
& 34.29 & 123.30 
\\
Demi-Stripped w/o Annotation   & 58.04 & 8.27 & 23.36 & 14.25 & 2.12
& 30.92 & 146.90 
\\
All-Stripped    & 62.14 & 9.80 & 25.11 & 16.88 & 2.51  
& 35.27 & 131.65 
\\
All-Stripped w/o Annotation    & 54.98 & 7.82 & 22.31 & 13.36 & 2.16   
& 32.40 & 150.75 \\ \hline
\multicolumn{8}{l}{\footnotesize * AVG time (function) is measured in seconds.}\\
\multicolumn{8}{l}{\footnotesize * AVG summary length is measured in words.}
\end{tabular}%
}
\label{table: Strip annotaiton}
\end{table*}

Since each phase uses a different dataset, their effectiveness should be verified independently rather than by direct score comparison.
Considering there is no universally accepted threshold for model evaluation scores, we provide the performance of another state-of-the-art binary code summarization model (CP-BCS~\cite{ye-etal-2023-cp}) as a baseline.
CP-BCS is tested on its own dataset, which consists of binaries compiled with GCC 7.3.0 for x86 architecture (32-bit) and then stripped.
Our results in both phases demonstrate higher usability compared to one of the existing method CP-BCS.

Also, during the training process of transfer learning, we try different training data proportions and select the optimal proportion, as shown in Table~\ref{table: performance in diff_prop}.

\begin{table}[htbp]
\caption{Performance in different training data proportion}
\resizebox{\columnwidth}{!}{%
\begin{tabular}{l|llll}
Proportion  & BLEURT-sum & BLEU & ROUGE-L & METEOR  \\ \hline
1:1        & 59.12      & 8.98  & 23.04   & 16.14 \\
1:2        & 59.81      & 9.15  & 23.50   & 16.42 \\
3:4        & 59.53      & 8.68  & 22.95   & 16.39 \\
4:3        & \textbf{62.14}     & \textbf{9.80}  & \textbf{25.11} & \textbf{16.88} \\ 
\hline
\multicolumn{5}{c}{\footnotesize * Limited by MalS, proportion 2:1 is unavailable.} \\
\multicolumn{5}{c}{\footnotesize * AVG summary length is measured in words.}
\end{tabular}%
}
\label{table: performance in diff_prop}
\vspace{-0.2cm}
\end{table}

After completing the training step, we use \textit{MalP} as the test set (it is the only pseudocode summary dataset built for malware that we know until now).
We compare \sysname to existing pseudocode summarization methods and prompt-based general-purpose large model on dataset \textit{MalP}. Simultaneously, we built three versions of data for \textit{MalP}: not-stripped, demi-stripped (only stripping the function name), and all-stripped (stripping all identifiers inside the function).

The closest version of the dataset to the real world, which is the results of the all-stripped version are shown in Table~\ref{table: comparision}. We conduct experiments using BLEURT-sum, BLEU, ROUGE, and METEOR (the usability of BLEURT-sum is substantiated in the referenced paper) to carry out experiments on \sysname, WizardLM~\cite{xu2024wizardlm}, and Code Llama~\cite{rozière2024code}, etc.
We also measure model performance using criteria such as summary length, variance, and processing time, as these factors can directly impact algorithm-based evaluations.

\begin{figure}[!htbp]
\centering
\captionsetup[subfloat]{font=scriptsize}
\subfloat[Malsight vs Code Llama-7b]{  
    \includegraphics[width=0.24\textwidth]{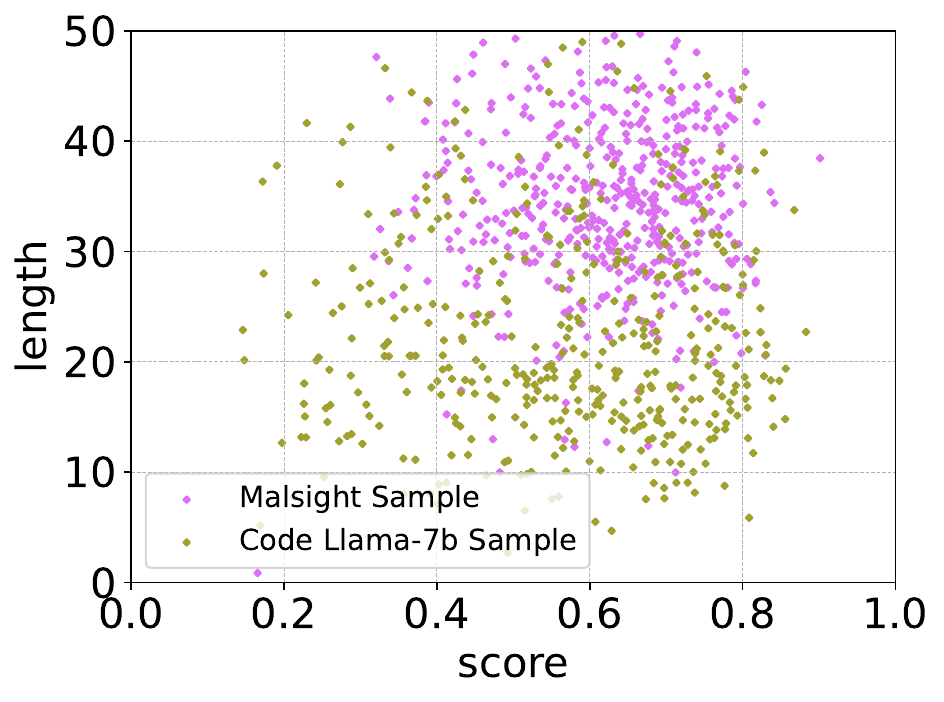}}
\subfloat[Malsight vs WizardCoder-15b]{  
    \includegraphics[width=0.24\textwidth]{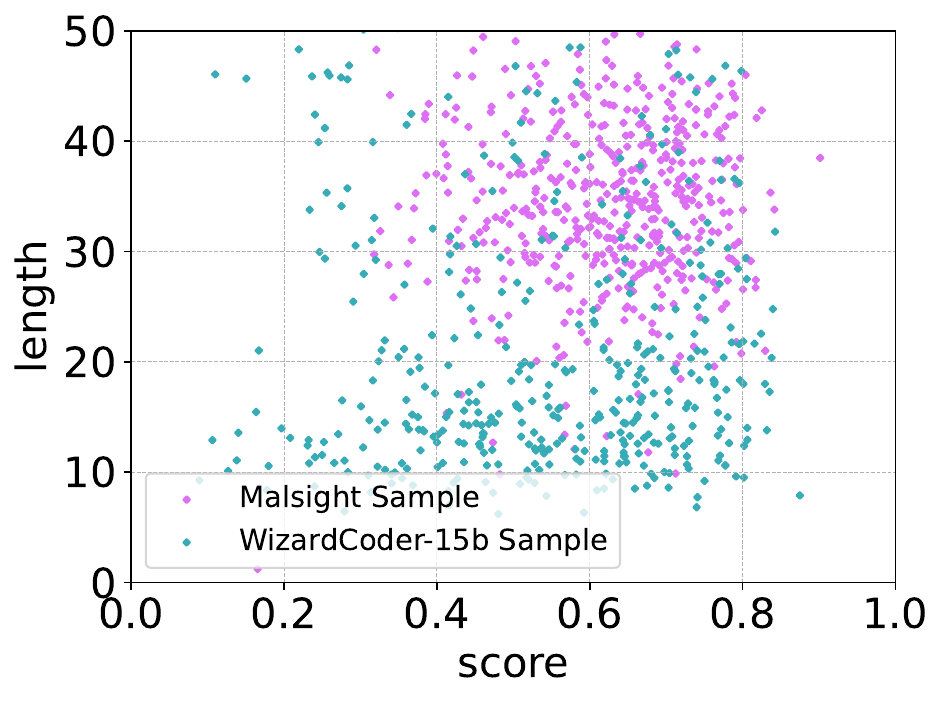}}
\caption{Data distribution. The segmentation distribution yielded by the proposed \sysname is more stable than the ones of the previous code-Llama.} 
\label{fig: Data distribution}
\vspace{-0.1cm}
\end{figure}
\vspace{-1mm}

General-purpose large models, trained on extensive code corpora, offer an advantage in code summarization tasks, with Code-Llama performing the best, achieving a BLEURT-sum score of 56.00.
\sysname's innovative fine-tuning approach offers a significant advantage in summarizing malware code, with its annotation generation effectively bridging the knowledge gap between specialized and general-purpose models. 
It has been revealed that \sysname achieved the highest scores among all evaluation indexes.

This conclusion can be corroborated by Fig.~\ref{fig: Data distribution}.
The figure shows that \sysname has better score evaluation results and less variance, i.e., more stable code summary output, on test results stripped of data.


\subsubsection{Annotation Extraction Effect}
The annotation extraction model is trained and tested on the \textit{AnnoS} dataset, shown in Table~\ref{table: Annotation extraction confusion matrix}.
The annotated extraction model is trained and tested on the \textit{AnnoS} dataset to complete the sequence SL task, dividing the data into normal codes (represented by N-label), important function APIs (represented by A-label), and important strings (represented by S-label).
The model test produces the confusion matrix shown in Table~\ref{table: Annotation extraction confusion matrix} below.

\begin{table}[htbp]
\caption{Annotation extraction confusion matrix}
\resizebox{\columnwidth}{!}{%
\begin{tabular}{l|lll}
Actual/Predicted & N-label(Predicted) & A-label(Predicted) & I-label(Predicted) \\ \hline
N-label(Actual)  & 1,847,647                  & 9,345                  & 11,721             \\
A-label(Actual)  & 15,865                  & 116,099                  & 1,784               \\
I-label(Actual)  & 21,639                  & 1,688                  & 38,933      \\ \hline           
\end{tabular}%
}
\label{table: Annotation extraction confusion matrix}
\end{table}

The results show that the accuracy of the model is 96.99\%, which is comparable to the results obtained by manual annotation.
Inevitably, some functions have ambiguous function names, or strings that appear to have certain characteristics, leading to negligible incorrect labeling.

\vspace{-0.1cm}
\subsection{RQ2. Ablation Study} \label{sec: RQ2}
\vspace{-0.1cm}

To verify the usability of each module in \sysname, we conduct ablation experiments on the annotation module and the two phases of code summarization. 
Further, We use \textit{MalP} as a test dataset, experimenting with different module combinations. 

\begin{table}[htbp]
\caption{Ablation experiment}
\resizebox{\columnwidth}{!}{%
\begin{tabular}{l|llll}
Model         & BLEURT-sum & BLEU & ROUGE-L & METEOR \\ \hline
Full Model    & \textbf{66.29}  & \textbf{10.95}  & \textbf{26.97}  & \textbf{18.80} \\
w/o Annotation    & 60.05     & 7.65    & 22.76     & 14.93     \\
w/o Phase 1   & 62.05        & 10.18   & 24.16     & 17.87     \\
w/o Phase 1 and Annotation  & 56.48    & 7.35    & 21.99     & 13.99   \\
w/o Phase 2   & 61.93        & 9.19    & 26.25     & 15.37      \\
w/o Phase 2 and Annotation  & 59.13    & 7.42    & 26.72     & 13.32     \\ \hline
\end{tabular}%
}
\label{table: Ablation}
\end{table}

\subsubsection{Module Ablation}
As shown in Table~\ref{table: Ablation}, the modules in \sysname complement each other effectively. 
This ablation study is conducted on the not-stripped version of \textit{MalP} to minimize result fluctuations due to the absence of dynamic annotation.

We conduct experiments under five settings: (1) Removing the annotation module and summarizing without any annotation, labelled as ``w/o Annotation''; (2) Canceling phase 1, labelled as ``w/o Phase ''; (3) Canceling phase 1 and removing the annotation module, labelled as ``w/o Phase 1 and Annotation''; (4) Canceling the phase 2, labelled as ``w/o Phase 2''; (5) Canceling the phase 2 and removing the annotation module, labelled as ``w/o Phase 2 and Annotation''.

The results indicate that different degrees of ablation have varying negative impacts on systematic performance.
``w/o Phase 1 and Annotation'' suffered a data shift and thus exhibited worse performance on the malware dataset than the other combinations, 56.48 in BLEURT-sum.
Notably, the absence of the annotation generation module (static annotation generation module) significantly affects performance, demonstrating its effectiveness.

\subsubsection{Annotater vs. Stripping}
The influence of different levels of the strip on the annotation module continues to be explored, and the results are shown in Table~\ref{table: Strip annotaiton}.
Experiments have shown that a higher degree of stripping creates a tolerable performance degradation in the absence of annotation, demonstrating the robustness of \sysname.

After the annotation module is ablated, it produces a 9.41\% performance degradation (measured by a BLEURT-sum score) in the not-stripped test and nearly 12\% in the all-stripped test, which has proved that the annotator effectively resists the adverse conditions caused by stripping.

\begin{figure}[htbp]
\centering
\includegraphics[width=0.5\textwidth]{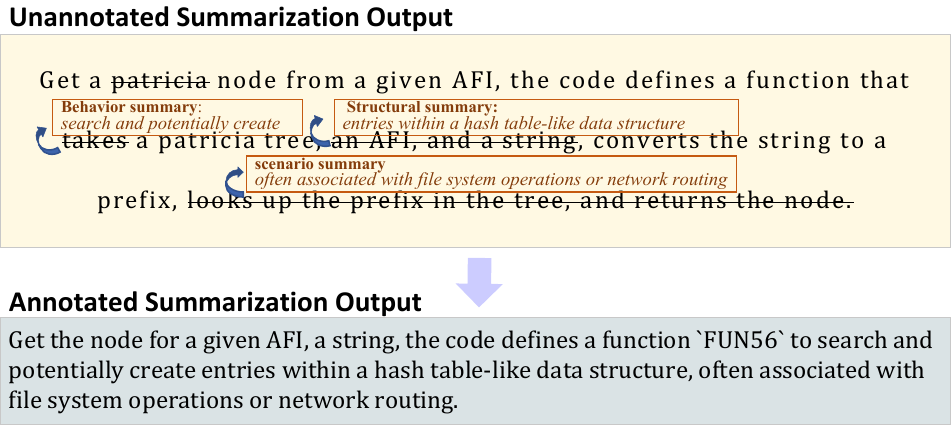}
\caption{Annotation ablation. The annotation generation module enriches sentence A with semantic information, enabling it to convey function functionality, which is lost after ablation.} 
\label{fig: Annotation Ablation}
\vspace{-0.3cm}
\end{figure}

In the generated results, we select relatively representative sentences, as shown in Fig.~\ref{fig: Annotation Ablation}.
In the yellow box, the output from the generation module without annotations shows that the large model is highly susceptible to biases due to hallucination~\cite{10.1145/3571730}.
Based on our observations, hallucinations can cause the code summary model to make false guesses about the types of parameters and internally called functions, leading to potential user misdirection.
To mitigate this issue, we incorporate annotations, which not only enhance the model's ability to summarize behavior, structure, and application scenarios but also reduce the nonsensical outputs caused by these illusions.

\vspace{-2mm}
\subsection{RQ3. Evaluation Algorithm Test} \label{sec: RQ3}

\begin{figure}[htbp]
\centering
\captionsetup[subfloat]{font=scriptsize}
\subfloat[BLEU]{  
    \label{fig: bleu}
	\includegraphics[width=0.24\textwidth]{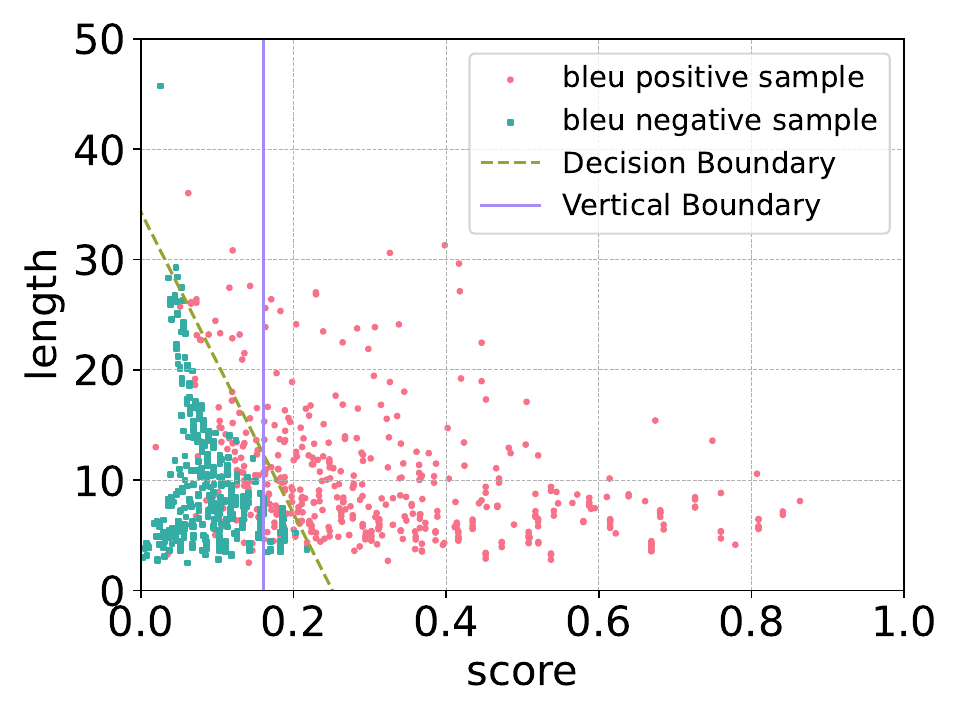}}
\subfloat[METEOR]{
	\includegraphics[width=0.24\textwidth]{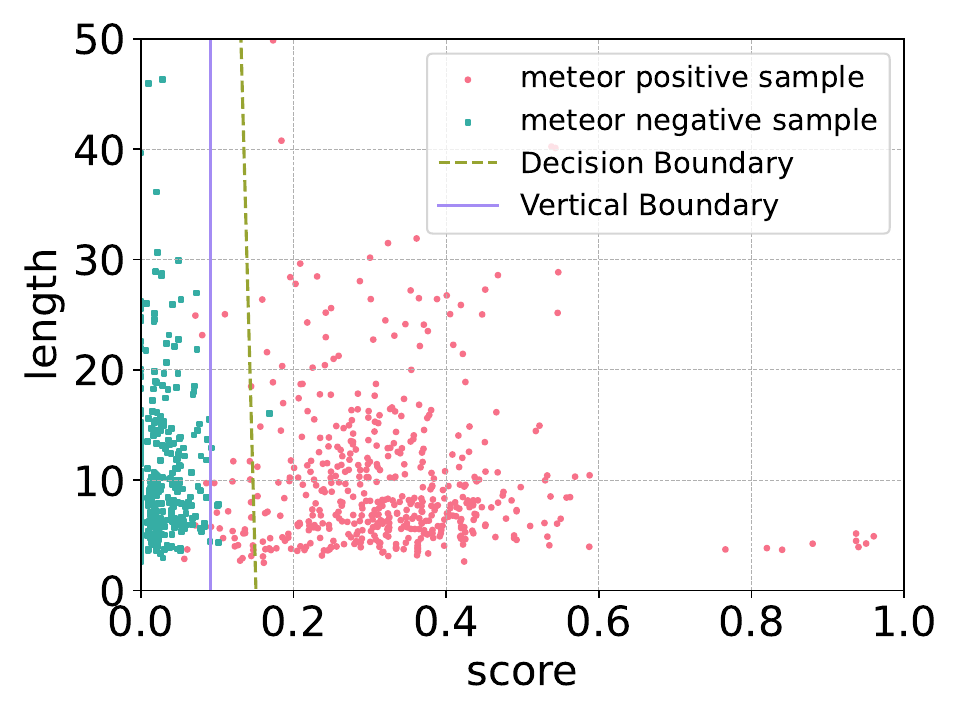}}

\subfloat[Word2vec]{
	\includegraphics[width=0.24\textwidth]{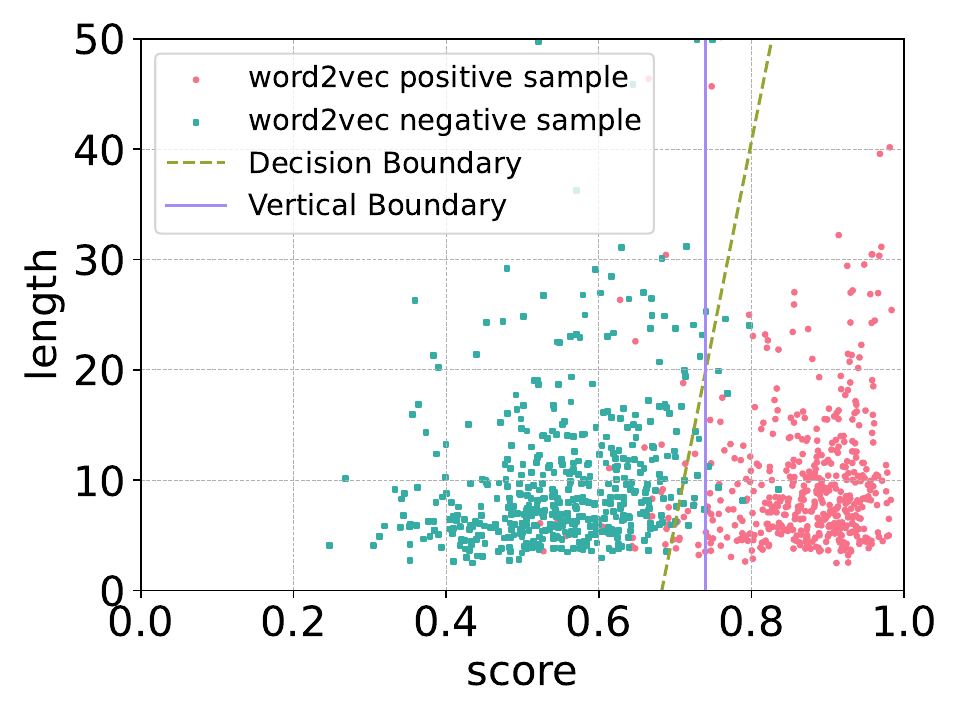}}
\subfloat[MoverScore]{
	\includegraphics[width=0.24\textwidth]{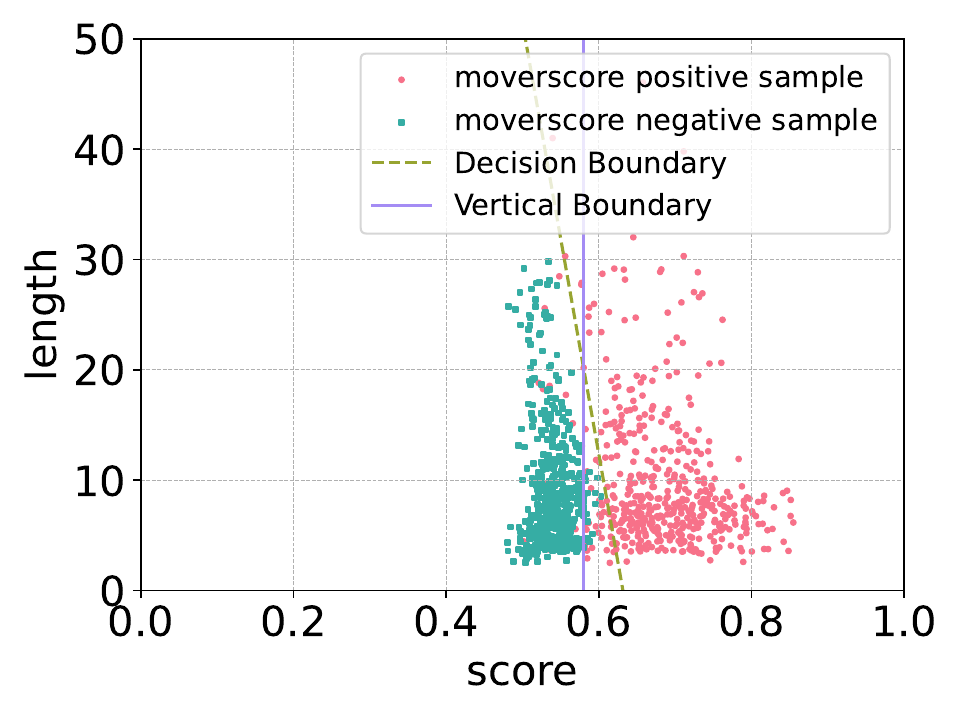}}
\caption{Test results of popular text similarity algorithms. The evaluation effect of two common algorithms and ML-based evaluation methods is not ideal, and their positive samples and negative samples show a certain degree of crossover.}
\label{fig: 4 evaluation alg}
\vspace{-1mm}
\end{figure}

As previously mentioned, we use positive and negative samples to test the effectiveness of these evaluation methods by assessing their ability to distinguish between the two. 
The performance of existing methods, illustrated in Fig.~\ref{fig: 4 evaluation alg}, shows varying degrees of crossover between positive and negative samples for BLEU, METEOR, Word2vec, and MoverScore. 
These four methods represent the current approaches in words' overlap measure(BLEU and METEOR) and words' embedding measure(word2vec and MoverScore), respectively.

\begin{figure}[htbp]
\centering
\captionsetup[subfloat]{font=scriptsize}
\subfloat[BLEURT-sum]{  
	\includegraphics[width=0.24\textwidth]{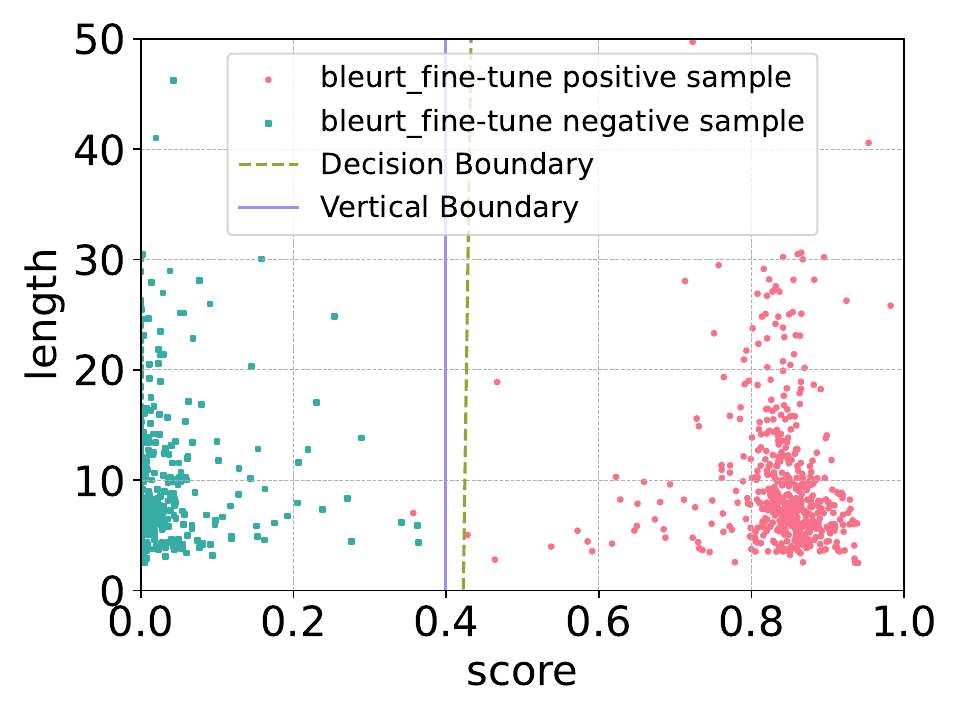}}
\subfloat[ROC Curve (Scaled)]{
	\includegraphics[width=0.24\textwidth]{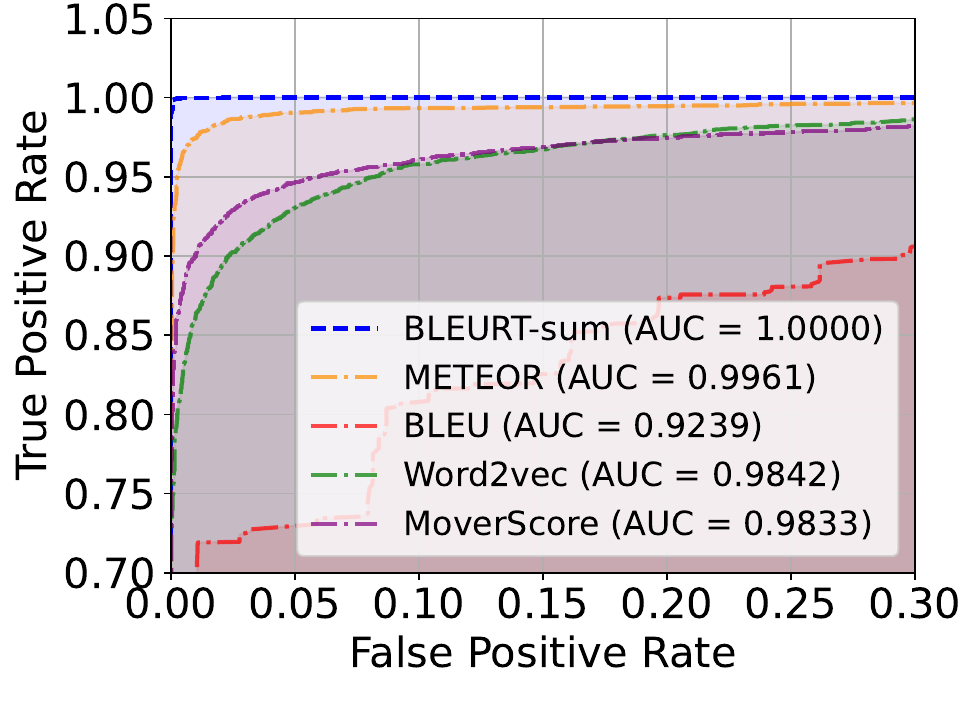}}
\caption{BLEURT-sum performance. BLEURT-sum has a significantly better ability to distinguish between positive and negative samples(a). Simultaneously, the ROC curve demonstrates a commendable AUC.(b)}
\label{fig: bleurt-sum eva}
\vspace{-1mm}
\end{figure}

In Fig.~\ref{fig: 4 evaluation alg}(a), we have established two types of decision boundaries: one being orthogonal to the X-axis (i.e., distinguishing between positive and negative samples by setting a threshold), and the other being a linear function forming a slanted line.
The same process is represented by BLEURT-sum as shown in Fig.~\ref{fig: bleurt-sum eva}, which shows the superiority of our method.

Specifically, we calculate the F1-score for these six methods (ROUGE-L is not shown in Fig.~\ref{fig: 4 evaluation alg}) on the positive and negative sample classification task of code summary sentences to measure their ability.
Our method achieves an F1-score exceeding 0.9999, significantly outperforming all other evaluation methods. 
Among the existing methods, METEOR performs best with an F1-score of 0.9811, while BLEU gets the lowest performance at only 0.85. None of the other techniques achieve an F1-score above 0.95.

Notably, our dataset is not meticulously curated, the negative samples are entirely random sentences with different meanings, which differ from real-world scenarios where the differences in meaning might be subtler.
An F1-score below 95\% indicates a tendency to misclassify, with approximately 1 in 10 samples judged incorrectly.
As an evaluation method, the misjudgment can be further amplified by the model results evaluated using the evaluation scheme in the following works.


\subsection{RQ4. Real World Experiments} \label{sec: RQ4}
\subsubsection{Human Evaluation Experiments}
So far, we have evaluated \sysname's performance at the function level, but this hasn't fully demonstrated the dynamic annotation module's capabilities.
Since real-world malware predominantly appears as executables, we manually analyzed ten malware samples and selected \textbf{79} critical functions. 
Three experienced reverse engineers collaboratively provided summaries for these functions.

\sysname's complete workflow is applied to these samples, with outputs evaluated using both BLEURT-sum and human evaluation metrics. 
We also calculate the variance deviation to assess the stability of the solution and measure the time required for summarizing a single malware sample.

For human evaluation, we invite ten evaluators, including five expert reverse engineers, three experienced engineers, and two beginners. 
They rate the summaries based on usability with scores of 0, 0.5, or 1, corresponding to unusable, partially usable, and usable, respectively. 
Given the subjectivity of human evaluation, we focus solely on usability rather than providing more detailed scores.

\subsubsection{Performance From Different Evaluators}

After three sets of scores, we obtain the human assessment scores shown in Table~\ref{table: three human}.
Experienced evaluators tend to give more conservative and stable scores, while less experienced evaluators provide a wider range of scores with higher variance. Further, we take the arithmetic average of the scores from the evaluators as the final human assessment score.

\begin{table}[htbp]
\caption{Results of Different Evaluators}
\resizebox{\columnwidth}{!}{%
\begin{tabular}{c|ccc}
     & Rich in experience & Experienced & Beginner \\ \hline
    Score & 58.14  & 59.51 & 64.73 \\
Variance & 173.12  & 197.49 & 237.44  \\
\hline
\end{tabular}%
}
\label{table: three human}
\vspace{-0.06cm}
\end{table}

Table~\ref{table: Real Malware Performance} shows the mean value of BLEURT-sum and the mean value of human evaluation on 79 labeled functions. The human evaluation score of 59.87 means that most functions have exceeded the partially usable standard.
\begin{table}[htbp]
\caption{Real Malware Performance}
\resizebox{\columnwidth}{!}{%
\begin{tabular}{c|ccc}
     & BLEURT-sum & Human Evaluation Score& AVG time (file) \\ \hline
    Score & 47.22 & 59.87  &1.90 \\
Variance & 161.61  & 193.34 & 1.21       \\ 
\hline
\multicolumn{4}{l}{\footnotesize * AVG time (per file) is measured in hours.}
\end{tabular}%
}
\label{table: Real Malware Performance}
\vspace{-0.2cm}
\end{table}

\begin{figure}[htbp]
\vspace{-0.2cm}
\centering
\captionsetup[subfloat]{font=scriptsize}
\subfloat[Scatter Plot]{  
    \includegraphics[width=0.24\textwidth]{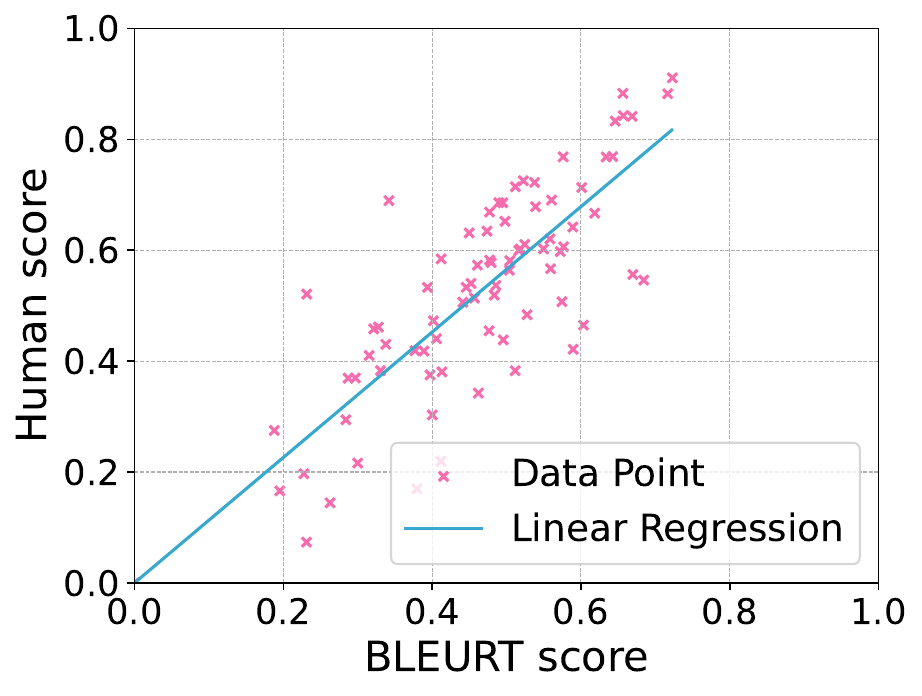}}
\subfloat[Violin Plot]{  
    \includegraphics[width=0.24\textwidth]{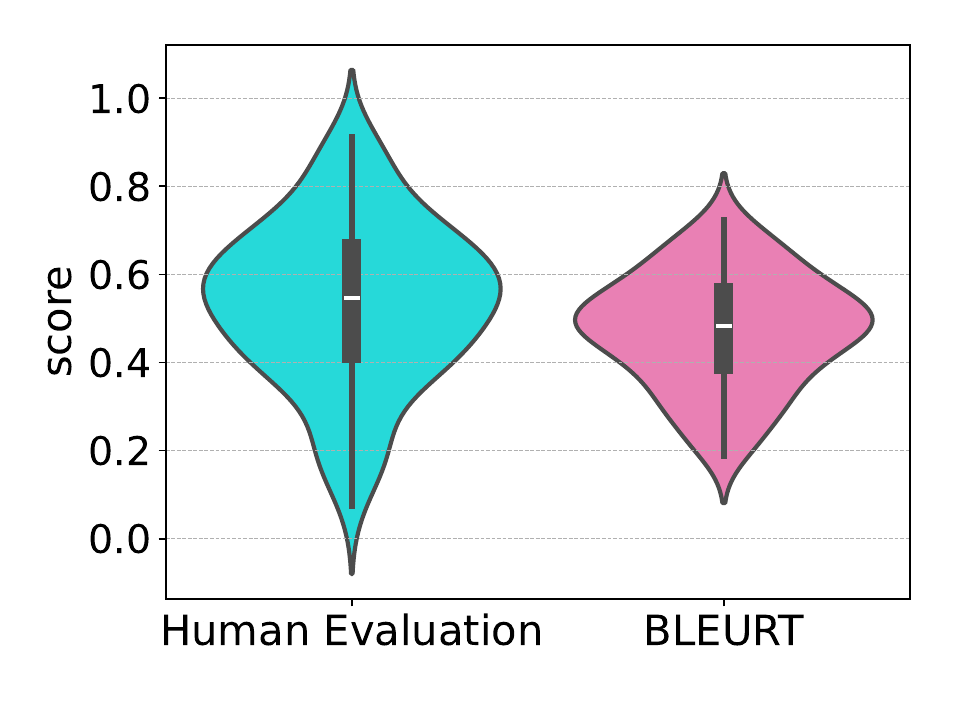}}
\caption{Real-world Performance. While the ratio of human assessment score to BLEURT-sum remained positive, \sysname can obtain higher assessment results.}
\label{fig: real-world performance}
\end{figure}

Further, we form a data point by combining a BLEURT-sum score with a human evaluation score for the same function's code summary. Thus, we plot a distribution of 79 data points in Fig.~\ref{fig: real-world performance}(a). 
It indicates that the BLEURT-sum has a positive linear correlation with human evaluation metrics, which shows its rationality. 
Furthermore, the data distribution in Fig.~\ref{fig: real-world performance}(b) indicates that, across the two distinct evaluation metrics, the majority of the evaluation scores for \sysname's outputs are concentrated within the middle to high range, which shows that our scheme is also usable in the real world.
\section{Case Study}

In Section RQ4, we evaluate \sysname's performance on ten real-world malware samples. 
To provide deeper insights into \sysname's varying performance across functions, we present two functions to illustrate the process behind generating a high-quality output and a lower-quality one.

\begin{figure}[!htbp]
\centering
\includegraphics[width=0.5\textwidth]{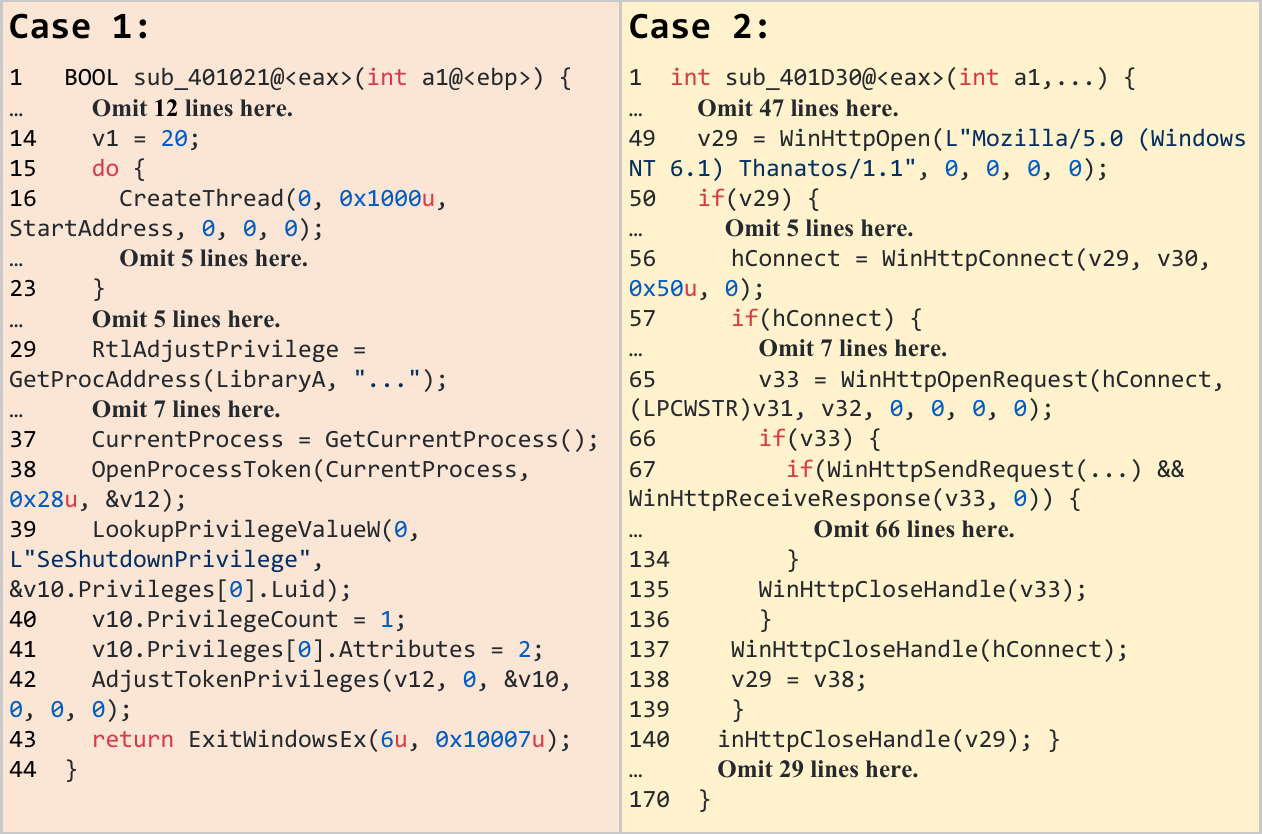}
\caption{Pseudocode of case 1\&2. The two real malware functions produced varying results, with the model's performance slightly weaker on the longer function due to length limitations.} 
\label{fig: pseudocode}
\vspace{-0.3cm}
\end{figure}

\noindent
\textbf{Case 1: sub\_401021.}
\begin{itemize}
    \item \sysname Summary: Set privilege, the code snippet is difficult to ascertain the function's purpose, but it likely retrieves or obtains professional information or status.
    \item Human Summary: Escalating the privileges of the virus process.
\end{itemize}

\sysname 's summary shown in Fig.~\ref{fig: pseudocode}(right) is considered \textbf{high-quality}.
In this case, \sysname summary accurately identifies ``setting privileges'' as the key point from a lengthy function, which aligns with the task of ``privilege escalation'' in the human summary.


\noindent
\textbf{Case 2: sub\_401D30.}
\begin{itemize}
    \item \sysname Summary: Convert a string to a wide character, the code defines a function that performs string operations, handles exceptional conditions, and creates new objects, potentially Get the server name, the code defines a function that opens a WinHttp connection, connects to a server, and handles error messages and string operations Get the next character from the string, the code defines a function sub\_401470 that performs string operations, handles exceptional conditions, and Delete the server.
    \item Human Summary: Connect to a malicious domain and make a request to retrieve data using HTTP.
\end{itemize}

\sysname 's summary of the function shown in Fig.~\ref{fig: pseudocode}(left) is considered \textbf{lower-quality}. 
The summary correctly identifies the function's use of an HTTP connection to a server and data operations, aligning with the human summary. 
However, it misses that the server is malicious due to lacking background information. Additionally, the model's output tends to be lengthy and repetitive due to the function's large code body.
We believe this is related to the model's parameter size and the input length it can process, using a model with a larger parameter count could further improve \sysname's performance.



\vspace{-0.15cm}
\section{Discussion}


In this section, we delve into practical issues based on our analysis of the experimental results. Additionally, we explore aspects not covered in our work and propose potential solutions.

\textbf{Real-World Malware vs. Function Summaries:}
In our evaluation of real-world malware, the BLEURT-sum score for code summaries was approximately 30\% lower than the experimental score from \textit{MalP} in Section~\ref{sec: RQ1}. 
This discrepancy may arise from the model's difficulty in summarizing longer functions, which are more common in real malware. 
Our analysis reveals that 15.8\% of the malware functions exceed 1000 tokens, while most training set functions are under 300 tokens. 
This length introduces excessive information and noise, complicating the extraction of critical code fragments.

Therefore, we investigate code summaries at the code fragment level, discovering that the model offers enhanced insights in this format. 
We propose adopting a large model agent approach for generating summaries of code fragments, which can then be integrated to form function-level summaries. 
Furthermore, utilizing additional labelled data from manual dynamic debugging can further enhance the results.



\textbf{Summary for assembly language rather than pseudocode:}
Efforts have been made to recover information from assembly code and to generate function summaries~\cite{10.1145/3460120.3484587}. 
A common perspective is that decompiled assembly code suffers varying degrees of information loss during the generation of pseudocode as Intermediate Representation (IR), depending on the disassembly algorithm used. 
Thus, starting directly from assembly language is considered a viable solution.

In our comparison of assembly code and pseudocode representations, we find that assembly code is challenging for models pre-trained on high-level languages to understand~\cite{tan2024llm4decompile}. 
The structural features of assembly code are completely different from those of high-level languages. 
Fine-tuning a model using pseudocode can confuse the model's understanding of function-level structural features, leading to unacceptable error output, possibly due to insufficient datasets.
On the other hand, the presence of lengthy assembly code can significantly increase the inference overhead for code summarization models. 
Moreover, the substantial differences in assembly code produced by various compilers may introduce data shifts. 
This issue is mitigated by consistently treating all code as pseudocode during processing.

\section{Conclusion}
In this paper, we addressed the challenges of binary malware summarization by introducing \sysname, a novel framework that leverages insights from both malicious source code and benign pseudocode. 
\sysname was designed to overcome the limitations of existing methods in handling entangled logic and stripped semantics commonly found in malware binaries. Our contributions include the development of two key datasets, MalS and MalP, specifically constructed to facilitate malware summarization. 
We also introduced MalT5, a lightweight LLM-based model tailored for this task, and proposed BLEURT-sum, an innovative evaluation metric that enhances code summarization performance beyond traditional word overlap metrics. Experimental results on both test sets and real-world malware datasets demonstrated the effectiveness of \sysname compared to other state-of-the-art methods. The framework successfully addressed critical issues such as recovering meaningful summaries from complex and stripped pseudocode, underscoring its robustness in handling real-world challenges. 

Future research may focus on improving the model's ability to generate concise summaries for longer code sequences and applying the \sysname framework to more complex downstream tasks. Additionally, extending the approach to further real-world malware scenarios remains a crucial direction for continued development.

\appendices

\section{Analysis of Traditional Similarity Evaluation Metric}

\textbf{Problems caused by BLEU algorithm: }
The formula BLEU uses to calculate sentence similarity is shown in equation~\ref{equ: bleu—score}:

\begin{equation}
BLEU=BP*exp(\sum\limits_{n=1}^{N}w_n\ln(Precision_n))
\label{equ: bleu—score}
\end{equation}

Where $BP$ is a brevity penalty factor, $w_n$ is the weight of the $n$-gram, and $Precision_n$ is the precision of the generated candidate sentence.
Specifically, equation~\ref{equ: bleu—score-Precision_n} shows the construction of $Precision_n$, where $candidate\&reference$ is the number of overlapping occurrences between the candidate sentence and the reference sentence.

\begin{equation}
Precisio n_n=\frac{len(candidate \& reference)+1}{len(candidate)+1}
\label{equ: bleu—score-Precision_n}
\end{equation}

In this context, the Add-One Smoothing method is used to avoid zero-count problems when the n-gram size is large. However, this introduces another issue: even if the candidate sentence and reference sentence are completely unrelated, this smoothing method still produces a certain score. This effect is particularly noticeable in the case of short sentences.

\begin{figure}[htbp]
\centering
\includegraphics[width=0.375\textwidth]{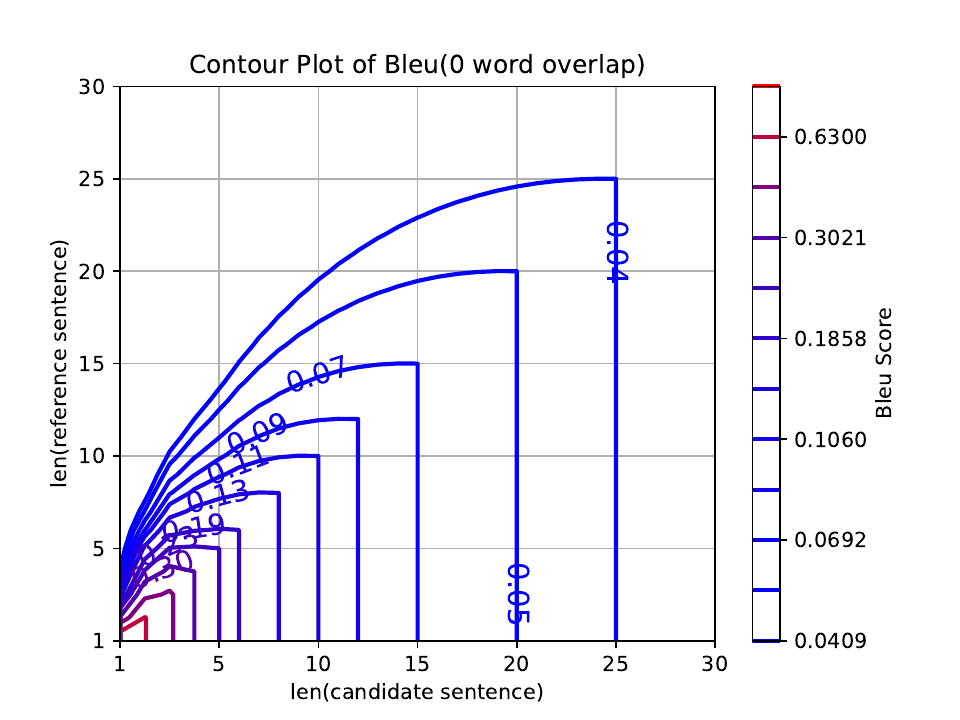}
\caption{Bleu Score When Zero Overlap. When two sentences have no overlap, due to the structural flaws in the BLEU algorithm, they still receive a score, indicating a bias toward shorter sentences.} 
\label{fig: bleu 0 lap}
\end{figure}

We conducted experimental tests for this scenario, testing each reference-candidate pair within the sentence length range of [1,30]. Each sentence pair had zero word overlap, and our results are shown in Figure \ref{fig: bleu 0 lap}.

Even when sentence pairs are completely mismatched, BLEU scores greater than 0.3 can occur for shorter sentences. This significant deviation from reality indicates that BLEU's scoring is distorted for short sentences in some cases.

\textbf{ROUGE \& METEOR, The flaw of calculating similarity in basic units of words:}
ROUGE and METEOR have something in common in the construction of sentence similarity evaluation algorithms. ROUGE follows the following equation~\ref{equ: rouge—score}.
\begin{equation}
ROUGE-L = F_{LCS} = \frac{(1+\beta^2)R_{LCS}P_{LCS}}{R_{LCS} + \beta^2P_{LCS}}
\label{equ: rouge—score}
\end{equation}

Where $R_{LCS}=\frac{LCS(C,S)}{len(S)}$,$P_{LCS}=\frac{LCS(C,S)}{len(C)}$ and $\beta$ is used to give weight to recall rates. $LCS(C, S)$ is used to calculate the length of the common substring of two strings $C$ and $S$.
Subjectively, when two target sentences have a higher degree of overlap in a specific word, they are given a higher ROUGE score.

METEOR designs on the basis of rouge, following equation~\ref{equ: meteor—score1}, \ref{equ: meteor—score2} and \ref{equ: meteor—score3}.
\begin{equation}
F_{mean}=\frac{(1+\beta^2)PR}{R+\beta P}
\label{equ: meteor—score1}
\end{equation}

\begin{equation}
Penalty=\gamma (\frac{chunks}{unigrams\_matched})^\theta
\label{equ: meteor—score2}
\end{equation}

\begin{equation}
METEOR=F_{mean}(1-Penalty)
\label{equ: meteor—score3}
\end{equation}

Where $P=\frac{n}{len(candidate)}$,$R=\frac{n}{len(reference)}$ and $n$ are the number of words where the candidate sentence and reference sentence overlap.
METEOR employs exact matching, stem matching, and WordNet-based synonym matching to address the issue of words with identical meanings not being recognized as overlaps. 
Consequently, METEOR outperformed ROUGE in our experiments, ranking second only to BLEURT-sum. However, algorithms that rely solely on word overlap can still misjudge due to structural similarities in sentences or phrases. 
For instance, in code summarization, sentences might share terms like "function", "aims", or "code," or convey the same idea using different wording, such as "compare two sentences" versus "bitwise and return true/false."

Interestingly, we found that expanding the stopword list appropriately can enhance the performance of word overlap-based methods, especially METEOR.

\begin{table*}[htbp]
\centering
\caption{Performance in different training data proportion}
\resizebox{\textwidth}{!}{%
\begin{tabular}{c|cccccc}
Proportion  & BLEURT-sum & BLEU & ROUGE-L & METEOR & AVG summary length & BLEURT-sum variance \\ \hline
1:1        & 59.12      & 8.98  & 23.04   & 16.14  & 39.83  & 181.01 \\
1:2        & 59.81      & 9.15  & 23.50   & 16.42  & 40.75  & 165.51 \\
3:4        & 59.53      & 8.68  & 22.95   & 16.39  & 42.21  & 173.39 \\
4:3        & \textbf{62.14}     & \textbf{9.80}  & \textbf{25.11} & \textbf{16.88} & 35.27 & \textbf{131.65} \\ 
\hline
\multicolumn{7}{l}{\footnotesize * Due to the limited size of MalS, the model cannot be fine-tuned in the training data proportion 2:1.} \\
\multicolumn{7}{l}{\footnotesize * AVG summary length is measured in words.}
\end{tabular}%
}
\label{table: performance in diff_props}
\vspace{-0.2cm}
\end{table*}

\begin{table*}[hbtp]
\centering
\caption{Summarization examples on TLS-related Go code by various LLMs.}
\label{tab:case_on_other_llms}
\resizebox{\textwidth}{!}{%
\begin{tabular}{l|p{0.875\textwidth}}
\toprule
\textbf{Method (LLM)} & \textbf{Summary} \\
\midrule
BinT5 & \textit{Go TLS connection handling with runtime checks.} \\
\midrule
HexT5 & \textit{TLS handshake timeout configuration.} \\
\midrule
WizardCoder-15B & \textit{This Go code sets up TLS handshake timeouts for HTTP servers, and it ensures secure connections by configuring read/write deadlines during the TLS handshake phase, preventing the server from hanging indefinitely. Additionally, it includes runtime checks to ensure compatibility.} \\
\midrule
Code Llama-7B & \textit{The code enforces TLS handshake timeouts to prevent hanging connections, setting deadlines during negotiation. It includes runtime checks to ensure compatibility while keeping encrypted chats from dozing off mid-conversation.} \\
\midrule
CodeT5+ & \textit{This code sets TLS timeouts for HTTP servers, ensuring security.} \\
\midrule
deepseek-llm-7b-chat & \textit{This code formats an integer, performs string operations, initializes TLS connection, handles time, and conditionally closes a connection or returns a Go interface.} \\
\bottomrule
\end{tabular}%
}
\vspace{-0.5cm}
\end{table*}

\section{Performance in different proportion of two-phase fine-tuning  training data size}
We sampled a total of 140,000 pieces of data in different proportions from the two datasets (MalS and BenignC) for two phases of fine-tuning. The results of the evaluation of model performance in each training data proportion are shown in the Table~\ref{table: performance in diff_props}.
In the experimental part of the text, we choose the ratio of 4:3 with the best effect as the actual training set ratio of \sysname.
Therefore, in the subsequent experiments, we adopt this configuration as the default training set composition for \sysname.
\begin{figure}[h]
\centering
\includegraphics[width=0.40\textwidth]{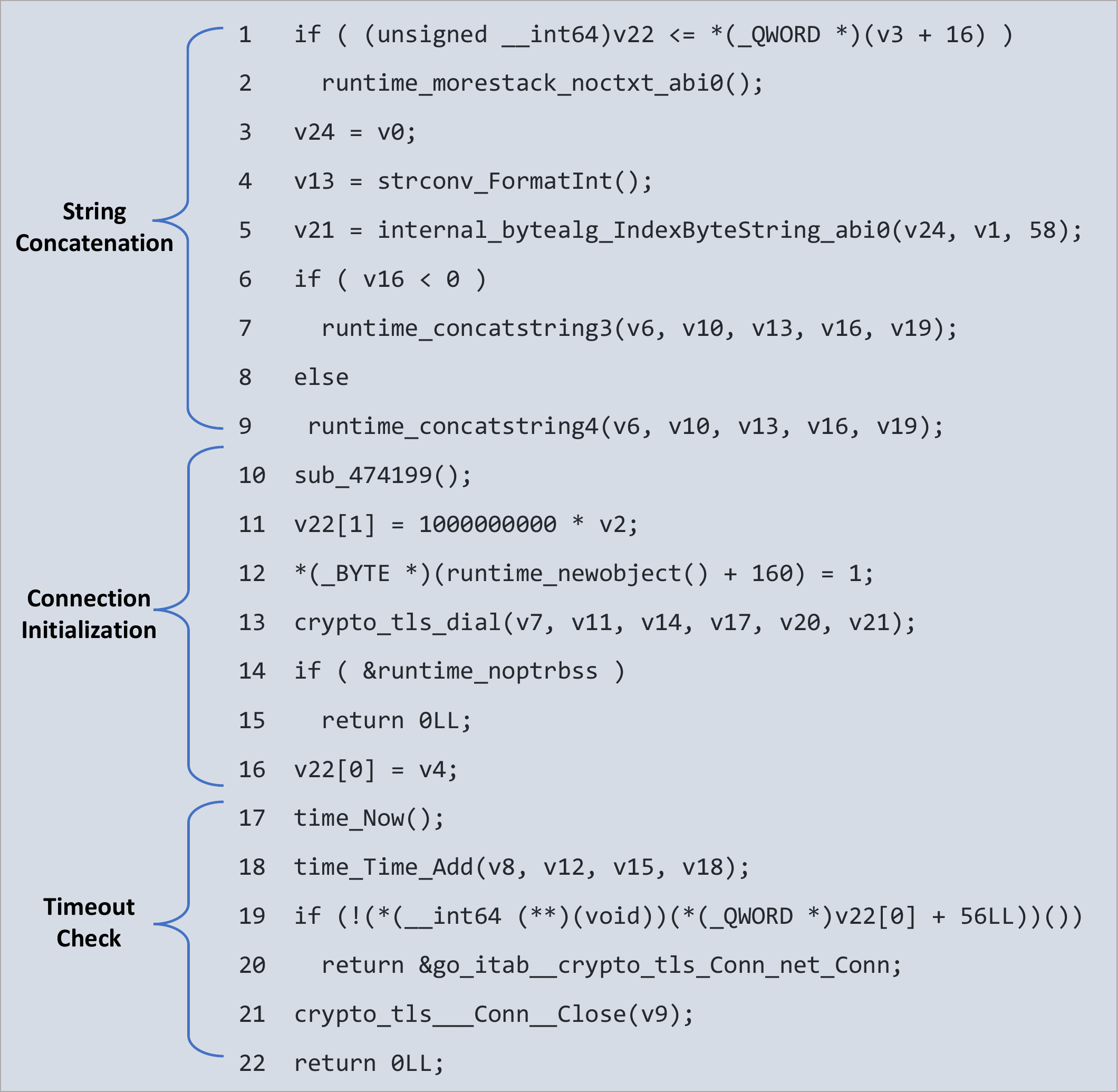}
\caption{Application of MALSIGHT to other binary code summarization tasks.} 
\label{fig: go-pscode}
\end{figure}

\section{Case Study Supplement}

\subsection{\sysname for other binary code summarization task}
We complement this with a case study on a pseudo-code summarization task on binary files written and compiled in \textbf{Go} (Golang) (shown in Fig.~\ref{fig: go-pscode}), aiming to demonstrate the generalizability of \sysname.

\noindent
\textbf{Case: Go Binary Psuedo-code.}
\begin{itemize}
    \item \sysname Summary: Initiate TLS connection, the code formats an integer into a string, performs various string operations including concatenation, initializes a TLS connection, handles time-related operations such as adding time, and conditionally closes the connection or returns a handle based on runtime checks.
\end{itemize}
\vspace{-0.4cm}
\subsection{Application of other LLMs to the binary summarzation task}

We would like to show the application of other large language models to the binary summarization task. We will use the same pseudo-code snippet (Fig.~\ref{fig: go-pscode}) shown in the previous subsection. The results are presented in Table~\ref{tab:case_on_other_llms}.

\let\oldthebibliography\thebibliography
\let\endoldthebibliography\endthebibliography
\renewenvironment{thebibliography}[1]{%
  \begin{oldthebibliography}{#1}%
  \setlength{\parskip}{0.0pt}%
  \setlength{\itemsep}{0.0em}%
}%
{%
  \end{oldthebibliography}%
}

\bibliographystyle{IEEEtran}
\bibliography{IEEEabrv, references}

\vfill

\end{document}